\documentclass[11pt,letterpaper]{article}

\usepackage{times}
\usepackage{url}
\usepackage{graphicx}
\usepackage{subfigure}
\usepackage{color}
\usepackage{epsfig}
\usepackage{latexsym}

\usepackage{amssymb}
\usepackage{amsthm}
\usepackage{framed}
\usepackage{a4wide}
\usepackage{enumerate}
\usepackage{makeidx}
\usepackage{bm}
\usepackage{epsfig}
\usepackage{graphics}
\usepackage{colortbl}
\usepackage{colordvi}
\usepackage{verbatim}
\usepackage{amsmath}
\usepackage{multirow} 
\usepackage{booktabs} 

\newcommand {\C} {{\rm I\kern-5.5pt C}}

\newcommand{\bP}[1]{{\mathbb{P}}\left[{#1}\right]}
\newcommand{\bE}[1]{{\mathbb{E}}\left[{#1}\right]}

\newcommand{\1}[1]{{\bf 1}\left[#1\right]}       

\newcommand{\fsquare}{\vrule height6pt width7pt depth1pt}   
\newcommand{\myproof}{{\hfill \\ \bf Proof. \ }}           
\newcommand{\myendpf}{\hfill\fsquare \\[0.1in]}             
\newcommand{\myvec}[1]{{\mbox{\boldmath{$#1$}}}}

\newtheorem{theorem}{Theorem}[section]

\newtheorem{lemma}[theorem]{Lemma}
\newtheorem{proposition}[theorem]{Proposition}

\begin{document}

\sloppy

\title{Node isolation in large homogeneous\\
         binary multiplicative attribute graph models}

\author{Sikai Qu and Armand M. Makowski\\
        Department of Electrical and  Computer Engineering,\\
        and Institute for Systems Research \\
        University of Maryland, College Park, MD 20742.\\
        Email: schuylerqu@gmail.com, armand@isr.umd.edu\\
}

\maketitle

\noindent
\hrulefill\\
The {\em multiplicative attribute graph} (MAG) model was introduced by Kim and Leskovec as a mathematically tractable
model  of certain classes of real-world networks. It is an instance of hidden graph models, and 
implements the plausible idea 
that network structure is collectively shaped by attributes individually associated with nodes.
These authors have studied several aspects of this model, including its connectivity, the existence of a giant component,
its diameter and the degree distribution. This was done in the asymptotic regime 
when the number of nodes and the number of node attributes both grow unboundedly large, the latter scaling with the former under
a natural admissibility condition. 
In the same setting, we explore the existence (or equivalently, absence) of isolated nodes, a property not discussed in the original paper.
The main result of the paper is a {\em zero-one} law for the absence of isolated nodes;
this zero-one law coincides with that obtained by Kim and Leskovec for graph connectivity (although under slightly weaker assumptions).
We prove these results by applying the method of first and second moments in a non-standard way 
to multiple sets of counting random variables associated with the number of isolated nodes.

\noindent
\hrulefill\\

\pagebreak

\section{Introduction}
\label{sec:Introduction}

The {\em multiplicative attribute graph} (MAG) model is a mathematically tractable network model recently introduced 
by Kim and Leskovec \cite{KimLeskovec, KimLeskovec_UAI2011}; it implements the plausible idea 
that network structure is collectively shaped by attributes individually associated with nodes.
MAG models are a special case of {\em hidden variable} models discussed in earlier literature where
each node is endowed with a set of intrinsic (\lq\lq hidden") attributes, e.g., authority, social success, wealth, etc.,
and the creation of a link between two nodes expresses a mutual \lq\lq benefit" based on their attributes, e.g. see references
\cite{BogunaPastor, CCDM, FIKMMU, SC, Sodeberg_2002, YoungSchneinerman} for examples.
Here we consider the {\em homogeneous binary} MAG model where the basic idea is implemented as follows:
With $n$ nodes in the network, the attributes are modeled as $\{0,1\}^L$-valued random variables (rvs)
$\myvec{A}(1), \ldots , \myvec{A}(n)$ which are assumed to be independent and identically distributed (i.i.d.).
{\em Conditionally} on $\myvec{A}(1), \ldots , \myvec{A}(n)$, edges are then created in a mutually independent manner with
\begin{equation}
\bP{ \begin{array}{c}
\mbox{An edge exists between } \\
\mbox{node $u$ and node $v$} \\
\end{array} \Big | \myvec{A}(1), \ldots , \myvec{A}(n) } = Q_L(\myvec{A}(u),\myvec{A}(v)),
\quad 
\begin{array}{c}
u \neq v \\
u,v =1, \ldots , n \\
\end{array}
\label{eq:Probability_p}
\end{equation}
for some Borel symmetric mapping $Q_L: \{0,1\}^L \times \{0,1\}^L \rightarrow [0,1]$ (whose form is to be specified shortly).

For each $u=1, \ldots , n$, the $L$  components $A_{1}(u), \ldots , A_{L}(u)$ of the attribute vector $\myvec{A}(u)$ 
for node $u$ are assumed to be i.i.d. $\{0,1\}$-valued Bernoulli rvs with
\[
\bP{ A_{\ell}(u) = 1 } = \mu(1) 
\quad \mbox{and} \quad \bP{ A_{\ell}(u) = 0 } = \mu(0),
\quad \ell =1, \ldots ., L
\]
for some $0 < \mu(1), \mu (0) < 1$ such that $\mu(0) + \mu (1) = 1 $. 
The homogeneous binary MAG model is then specified by taking
\begin{equation}
Q_L(\myvec{A}(u),\myvec{A}(v)) = \prod_{\ell=1}^L q(A_{\ell}(u) ,A_{\ell}(v)),
\quad 
\begin{array}{c}
u \neq v \\
u,v =1, \ldots , n \\
\end{array}
\label{eq:Probability_q}
\end{equation}
for some symmetric $2 \times 2$ matrix $\mathcal{Q} \equiv ( q(a,b) )$ (with $0 < q(a,b) < 1$, $a,b=0,1$ and $q(0,1) = q(1,0)$). 
Formal definitions and a complete construction are provided in Section \ref{sec:MAG_Models}.
A useful way of thinking about this MAG model, especially relevant  in the context of social networks, is to imagine
that each network participant or node, answers a set of $L$ binary (YES/NO) questions, 
e.g., Does node $u$ exercise regularly? Does node $u$ belong to a book club? etc.
Then, $A_{\ell}(u) = 1 $ (resp. $A_{\ell}(u) = 0 $) can be interpreted as a YES (resp. NO) answer to the $\ell^{th}$ 
question answered by node $u$.

In \cite{KimLeskovec} Kim and Leskovec studied several aspects of this model, including its connectivity, the existence of a giant component,
its diameter and the degree distribution. This was done in the asymptotic regime 
when the number $n$ of nodes and the number $L$ of attributes both grow unboundedly large, the latter scaling with the former under
the condition $L_n \sim \rho \ln  n$ for some $\rho > 0$ (in which case the scaling $n \rightarrow L_n$ is said to be $\rho$-admissible).
In the same setting
we explore the existence (or equivalently, absence) of isolated nodes in the MAG model, 
a property which was not discussed in the original paper \cite{KimLeskovec}.
The main result is a {\em zero-one} law for the absence of isolated nodes; it takes
a different form depending on whether $1 + \rho \ln  \mu(0) > 0$ or  $1 + \rho \ln  \mu(0) < 0$,
the appropriate version being recorded 
in Theorem \ref{thm:Zero-OneLaw+Part1}  and Theorem \ref{thm:Zero-OneLaw+Part2}, respectively.
These results have the same structure as the zero-one law for graph connectivity obtained by Kim and Leskovec
\cite[Thm. 4.2, p. 126]{KimLeskovec} but are given here under weaker conditions.
See Section \ref{sec:Results} for details.

An undirected graph contains no isolated nodes if it is connected, but the converse is clearly not true in general.
However, in many random graph models these two graph properties obey {\em identical} zero-one laws;
this is known to occur for Erd\H{o}s-R\'{e}nyi graphs \cite{ER1959,ER1960}, 
random geometric graphs \cite{GuptaKumar, PenroseBook}, 
random key graphs \cite{Rybarczyk2011,YaganMakowskiConnectivity} and random threshold graphs \cite{MakowskiYagan_RTGs_JSAC},
to mention a few examples.
While this is not universally valid as can be seen from $k$-out-$n$ random graphs \cite{FennerFrieze} 
(also called pairwise graphs in \cite{YaganMakowskiPairwiseIT}),
our results establish its validity for the MAG model in the limiting regime considered here. 

To prove Theorem \ref{thm:Zero-OneLaw+Part1}  and Theorem \ref{thm:Zero-OneLaw+Part2} we apply the method
of first and second moments to various {\em count variables} associated with  the number of isolated nodes in MAG graphs:
Traditionally this well-worn approach is carried out in terms of the rv $I_n(L)$ 
which counts the number of isolated nodes in the MAG model with $n$ nodes and $L$ attributes per node. 
It relies on the basic observation that $\bP{ I_n(L) = 0 }$ coincides with the probability that there are no isolated nodes in the graph,
and leverages the elementary inequalities
\begin{equation}
1 - \bE{ I_n (L) } \leq \bP{  I_n (L)  = 0 } \leq 1 - \frac{ \left ( \bE{ I_n (L) } \right )^2}{\bE{ I_n(L)^2} }. 
\end{equation}
See Section \ref{sec:Roadmap} for details. In principle a successful completion of this program
requires exploring the limiting behavior of the sequences
of moments $\{ \bE{ I_n (L_n) } , \ n=2,3, \ldots \}$ and $\{ \bE{ I_n (L_n)^2 } , \ n=2,3, \ldots \}$ under the appropriate conditions.

For MAG models this is easier said than done, and we must resort to an indirect (and much finer) analysis: 
While the method of first moment can be successfully used
on the rv $I_n(L)$ in a rather straightforward manner,
applying the second moment method to the same rv $I_n(L)$ is problematic due to the complicated expressions for the quantities involved.
Instead we introduce additional count variables, namely the rv $I^{(\ell)}_n(L)$ 
which tallies the number of isolated nodes (amongst the $I_n(L)$ isolated nodes)
who have answered YES to exactly $\ell$ of the $L$ questions with $\ell=0, 1, \ldots , L$. Key here is the observation that
\begin{equation}
\bP{ I_n(L) = 0 } \leq  \bP{ I^{(\ell)}_n(L) = 0 },
\quad \ell =0,1, \ldots ,L.
\label{eq:Bound}
\end{equation}
We now give a summary of how this tailor-made approach is implemented:

\begin{enumerate}

\item[(i)] We start the analysis in Section \ref{sec:EvaluatingMoments} by evaluating the first two moments of
these count variables; expressions are given for the first moments in Lemma \ref{lem:EvalFirstMoment}
and for the second moments in Lemma \ref{lem:EvalSecondMoment}
(with the evaluation being completed in Appendix \ref{app:ProofLemmaEvalSecondMoment}).

\item[(ii)] Theorem \ref{thm:Zero-OneLaw+Part1} is established in Section \ref{sec:ProofThmPart1} and
its proof is rather short:
We begin with auxiliary \lq\lq zero-infinity" laws for the first moments under conditions that mirror
the ones of Theorem \ref{thm:Zero-OneLaw+Part1}.
Proposition \ref{prop:Infinity-OneLaw+FirstMoment+Part1} deals with the first moments of $\{ I_n(L_n), \ n=1,2, \ldots \}$ 
and allows us to show $\lim_{n \rightarrow \infty} \bP{ I_n(L_n) = 0 } = 1$ 
by the method of first moment under the conditions for the one-law.
Proposition \ref{prop:Infinity-OneLaw+FirstMoment+Part1-ell=0} captures the behavior of 
the first moments $\{ \bE{ I^{(0)} _n(L_n)}, \ n=1,2, \ldots \}$,
and leads to the desired zero-law follows via (\ref{eq:Bound}) (with $\ell=0$) upon showing that $\lim_{n \rightarrow \infty} \bP{ I^{(0)}_n(L_n) = 0 } = 0 $ 
by the method of second moment.

\item[(iii)]  The proof of Theorem \ref{thm:Zero-OneLaw+Part2} is in the same vein but is a lot more involved;
its major components are presented in Section \ref{sec:CompleteProofThmPart2}:
Here, two auxiliary \lq\lq zero-infinity" laws for the first moments are needed that parallel Theorem \ref{thm:Zero-OneLaw+Part2}.
Proposition \ref{prop:Infinity-OneLaw+FirstMoment+Part2} deals with the first moments of the rvs $\{ I_n(L_n), \ n=1,2, \ldots \}$
(as did Proposition \ref{prop:Infinity-OneLaw+FirstMoment+Part1} under the assumptions of Theorem \ref{thm:Zero-OneLaw+Part1}).
The first moment behavior of the rvs $\{ I^{(\ell_n)} _n(L_n), \ n=1,2, \ldots \}$ is obtained for certain integer-valued sequences $n \rightarrow \ell_n$
associated with the scaling $n \rightarrow L_n$ under certain conditions. 
This result,  which is reported in  Proposition \ref{prop:COMPLEMENT_Infinity-OneLaw+FirstMoment+Part2},
is  established in Section \ref{sec:Proof+COMPLEMENT_Infinity-OneLaw+FirstMoment+Part2}.

\item[(iv)]   We give two different proofs to Proposition \ref{prop:Infinity-OneLaw+FirstMoment+Part2}.
The first one is presented in Section \ref{sec:Proof+Infinity-OneLaw+FirstMoment+Part2}, and makes uses of Stirling's approximation
to evaluate the asymptotic behavior of various combinatorial quantities.
The second proof of Proposition \ref{prop:Infinity-OneLaw+FirstMoment+Part2} is given
in Section \ref{app:ZeroLawPart2} and Section \ref{app:InfinityLawPart2}, 
and relies on a change of measure argument introduced in Section \ref{app:ChangeMeasure}. 
While this second  proof may be construed as less intuitive than the one provided in Section \ref{sec:Proof+Infinity-OneLaw+FirstMoment+Part2},
it has the advantage of giving a probabilistic interpretation to the conditions appearing in Theorem \ref{thm:Zero-OneLaw+Part2}.

\end{enumerate} 

A word on the notation and conventions in use: Unless specified otherwise,
all limiting statements, including asymptotic equivalences, are understood with $n$ going to infinity. 
The rvs under consideration are all 
defined on the same probability triple  $(\Omega, {\cal F}, \mathbb{P})$. 
The construction of a probability triple sufficiently large to carry all required rvs is standard,
and omitted in the interest of brevity.
All probabilistic statements are made with respect to the probability
measure $\mathbb{P}$, and we denote the corresponding expectation operator by $\mathbb{E}$. 
We abbreviate almost sure(ly) (under $\mathbb{P}$) by a.s.
If $E$ is a subset of $\Omega$, then $\1{E}$ is the indicator rv
of the set $E$ with the usual understanding that $\1{E}(\omega) = 1$ (resp. $\1{E}(\omega) = 0$) if $\omega \in E$
(resp. $\omega \notin E$). 
The symbol $\mathbb{N}$ (resp. $\mathbb{N}_0$)  denotes the set of non-negative (resp. positive) integers.
We view sequences as mappings defined on $\mathbb{N}_0$; the mapping itself is denoted by
bolding the symbol used for the generic element of the corresponding sequence.
Unless otherwise specified, all logarithms are natural logarithms with $\ln x$ denoting the natural logarithm of $x > 0$.

\section{Homogeneous (binary) MAG models}
\label{sec:MAG_Models}

The MAG  model is parametrized by a number of quantities, chief amongst
them the number $n$ of nodes present in the network
and the number $L$ of attributes associated with each node -- Both $n$ and $L$ are positive integers.
Nodes are labeled $u=1,2, \ldots$, while attributes are labeled $\ell =1,2, \ldots$.
Each of the $L$ attributes associated with
a node is assumed to be binary in nature with $1$ (resp. $0$) signifying that the attribute is present (resp. absent).
We conveniently organize these $L$ attributes into a vector element $\myvec{a}_L= (a_1, \ldots , a_L)$ of  $\{0,1\}^L$.

\subsection{The underlying rvs}
\label{subsec:BasicRVs}

The propensity of nodes to attach to each other is governed by their attributes in a way to be clarified shortly.
The probability triple $(\Omega, {\cal F}, \mathbb{P} )$ is assumed to carry two collections of rvs, namely the  collection
\[
\left \{ A, A_\ell, A_\ell (u), \ \ell =1,2, \ldots ; \ u=1,2, \ldots
\right \}
\]
and the triangular array
\[
\left \{ U(u,v), \ u=1,2, \ldots ; \ v =u+1, u+2, \ldots  \right \} .
\]
The following assumptions are enforced throughout:

\begin{enumerate}
\item[(i)] 
The collection
$\left \{ A, A_\ell, A_\ell (u), \ \ell =1,2, \ldots ; \ u=1,2, \ldots \right \}$
and the triangular array
$\left \{ U(u,v), \ u=1,2, \ldots ; \ v =u+1, u+2, \ldots \right \}$
are {\em mutually independent};

\item[(ii)]
The rvs
$\left \{ U(u,v), \ u=1,2, \ldots ; \ v =u+1, u+2, \ldots \right \}$
are  {\em i.i.d.} rvs, 
each of which is {\em uniformly} distributed on the interval $(0,1)$;
and 

\item[(iii)] The rvs
$\left \{ A, A_\ell, A_\ell (u), \ \ell =1,2, \ldots ; \ u=1,2, \ldots \right \}$
form a collection of {\em i.i.d.} $\{0,1\}$-valued rvs
with pmf $\myvec{\mu} = (\mu(0), \mu(1) )$ where
$\bP{ A = 0 } = \mu (0)$ and $\bP{ A = 1 } = \mu (1)$.
To avoid trivial situations of limited interest, we assume that both $\mu (0)$ and $\mu(1)$ are elements of the
{\em open} interval $(0,1)$ such that $\mu(0) + \mu(1) =1 $.
\end{enumerate} 
For each $L=1,2, \ldots $, we write
\[
\myvec{A}_L = ( A_1 , \ldots , A_L )
\quad \mbox{and} \quad
\myvec{A}_L (u) = ( A_1(u) , \ldots , A_L(u) ),
\quad u=1,2, \ldots .
\]
Under the enforced assumptions,  the $\{0,1\}^L$-valued rvs
$ \left \{ \myvec{A}_L, \myvec{A}_L(u) , \ u=1,2, \ldots \right \} $
are i.i.d. rvs, each with i.i.d. components distributed like the generic rv $A$.
We shall also have use for the partial sum rvs
\begin{equation}
S_L (u) = A_1(u) + \ldots + A_L(u),
\quad u=1, 2, \ldots 
\label{eq:CountAttributes}
\end{equation}
and
\begin{equation}
S_L = A_1 + \ldots + A_L .
\label{eq:CountAttributesGeneric}
\end{equation}
For each $\ell=1,, \ldots $, we shall say that node $u$ exhibits (resp. does not exhibit) the $\ell^{th}$ attribute if $A_\ell(u)=1$ (resp. $A_\ell(u)=0$).
In that terminology, the rv $S_L(u)$ then counts the number of attributes exhibited by node $u$ amongst the first $L$ attributes.\footnote{In terms 
of YES/NO answers to binary questions, $S_L(u)$ then counts the number of YES answers given by node $u$ to the $L$ first questions.}
Under the enforced assumptions, the rvs $\{ S_L (u), \ u=1,2, \ldots \}$ form a sequence of i.i.d. rvs, 
each being distributed according to the rv $S_L$ which is itself a Binomial rv ${\rm Bin}(L,\mu(1))$.

For notational reasons we find it convenient to augment the triangular array of uniform rvs into the larger collection
$\left \{ U(u,v), \ u,v=1,2, \ldots \right \} $ through  the definitions
\[
U (u,u) = 1
\quad \mbox{and} \quad
U(v,u) = U(u,v),
\quad 
\begin{array}{c}
v = u+1, \ldots \\
u=1,2, \ldots
\end{array}
\]

\subsection{Adjacency}
\label{subsec:MAG-Adjacency}

On the way to defining homogeneous binary MAGs, we introduce notions of {\em adjacency} between nodes based on their attributes.
To do so we start with  an $2 \times 2$ matrix $\mathcal{Q}$ given by
\[
\mathcal{Q} 
\equiv
( q(a,b) )
= \left (
\begin{array}{cc}
q(1,1) & q(1,0) \\
q(0,1) & q(0,0) \\
\end{array}
\right ).
\]
Throughout we assume the {\em symmetry} condition 
\begin{equation}
q(1,0) = q(0,1),
\label{eq:Symmetry}
\end{equation}
together with the {\em non-degeneracy} conditions
\begin{equation}
0 < q(a,b) < 1,
\quad a,b \in \{0,1\} .
\label{eq:Non-DegeneracyConditions}
\end{equation}

Fix $L=1,2, \ldots $. With this symmetric $2 \times 2$ matrix $\mathcal{Q}$ we associate 
a mapping  $Q_L : \{0,1\}^L \times \{0,1\}^L \rightarrow [0,1]$ given by
\begin{equation}
Q_L(  \myvec{a}_L,  \myvec{b}_L )
= \prod_{\ell=1}^L q(a_\ell, b_\ell),
\quad 
\myvec{a}_L,  \myvec{b}_L \in \{0,1\}^L .
\label{eq:Q_L}
\end{equation}
Interpretations for these quantities will be given shortly.
The enforced assumptions 
(\ref{eq:Symmetry})-(\ref{eq:Non-DegeneracyConditions}) on $\mathcal{Q}$ readily imply
\begin{equation}
Q_L(  \myvec{b}_L,  \myvec{a}_L )
=
Q_L(  \myvec{a}_L,  \myvec{b}_L ),
\quad \myvec{a}_L,  \myvec{b}_L  \in \{ 0,1\}^L 
\label{eq:SymmetryCondition}
\end{equation}
with
\begin{equation}
0  < Q_L(  \myvec{a}_L,  \myvec{b}_L ) < 1,
\quad \myvec{a}_L,  \myvec{b}_L \in \{ 0,1\}^L.
\label{eq:ProbabilityConditions}
\end{equation}

Pick two nodes $u,v =1,2, \ldots $. 
We say that node $u$ is  {\em $L$-adjacent} to node $v$, written $u \sim_L v$, if the condition
\begin{equation}
U(u,v) \leq Q_L ( \myvec{A}_L(u), \myvec{A}_L(v) )
\label{eq:L-Adjacency}
\end{equation}
holds, in which case an (undirected) edge from node $u$ to node $v$ is said to exist.
Obviously, $L$-adjacency is a binary relation on the set of all nodes.
Since $U(u,v) = U(v,u)$, it is plain from (\ref{eq:SymmetryCondition})
that node $u$ is $L$-adjacent to node $v$ if and only if
node $v$ is $L$-adjacent to node $u$ -- This allows us to say that nodes $u$ and $v$ are $L$-adjacent
without any risk of confusion. 
Node $u$ cannot be $L$-adjacent to itself because $U(u,u) = 1$ (by convention) and $Q_L ( \myvec{A}_L(u), \myvec{A}_L(u) ) < 1$ by
(\ref{eq:ProbabilityConditions}) --
In other words,  $L$-adjacency will not give rise to self-loops.

We encode $L$-adjacency  through the $\{0,1\}$-valued rvs
$ \left \{ \chi_L(u,v), \ u,v =1, 2, \ldots  \right \} $ given by
\begin{eqnarray}
\chi_L(u,v)
=  \1{ U(u,v) \leq Q_L ( \myvec{A}_L(u), \myvec{A}_L(v) ) },
\quad u,v =1,2, \ldots 
\label{eq:EdgeAssignmentVariables}
\end{eqnarray}
with $\chi_L(u,v)=1$ (resp. $\chi_L(u,v)=0$) 
corresponding to the existence (resp. absence) of an (undirected) edge between node $u$ and node $v$.
In view of earlier remarks,  the conditions
\begin{equation}
\chi_L(u,u) = 0
\quad \mbox{and} \quad
\chi_L(v,u) = \chi_L(u,v),
\quad u,v=1,2, \ldots
\label{eq:Chi+Conditions}
\end{equation}
are all satisfied.

\subsection{Defining the homogeneous binary MAG models}

Fix $n=1,2, \ldots$ and $L=1,2, \ldots $.
The homogeneous binary MAG over a set of $n$ nodes, labelled $1, \ldots, n$, with
each node having $L$ attributes, labelled $1, \ldots , L$, is defined as the random graph
$\mathbb{M}(n;L)$ whose edge set is determined through the rvs
$\left \{ \chi_L(u,v), \ u,v=1,2, \ldots, n \right \}$.
From (\ref{eq:Chi+Conditions}) it follows that edges in $\mathbb{M}(n;L)$ 
are undirected and that there are no self-loops, hence any realization of $\mathbb{M}(n;L)$  is a simple graph.
For simplicity we shall refer to this model as the MAG model.

This definition is equivalent to the one given by Kim and Leskovec \cite{KimLeskovec}.\footnote{Strictly speaking, the definition given above
is slightly more restrictive than the one proposed in \cite{KimLeskovec} as we have eliminated by {\em construction} the possibility of self-loops,
whereas such links are neglected  by Kim and Leskovec  as making no contributions in the limiting regime. 
See the discussion after Theorem 3.1 in \cite{KimLeskovec}.}
Indeed, with the help of Assumptions (i) and (ii), it is a simple matter to check from (\ref{eq:EdgeAssignmentVariables})
that  the rvs forming the triangular array
\[
\left \{
\chi_L(u,v), \ 
\begin{array}{c} 
u=1, \ldots , n \\
v=u+1, \ldots , n \\
\end{array}
\right \}
\]
are  {\em conditionally independent} given the i.i.d. attribute random vectors
$\{ \myvec{A}_L(u), \  u=1,2, \ldots , n \}$ with
\begin{eqnarray}
\lefteqn{
\bP{ \chi_L(u,v)= 1  | \myvec{A}_L(w) ,  \  w=1,2, \ldots , n } 
} & &
\nonumber \\
&=&
\bP{ U(u,v) \leq Q_L ( \myvec{A}_L(u), \myvec{A}_L(v) )  | \myvec{A}_L(w) ,  \  w=1,2, \ldots , n } 
\nonumber \\
&=& 
Q_L(  \myvec{A}_L(u),  \myvec{A}_L(v) )
\nonumber \\
&=& 
\prod_{\ell=1}^L q(A_\ell (u), A_\ell (v)),
\quad
\begin{array}{c} 
u \neq v \\
u,v=1, \ldots , n \\
\end{array}
\label{eq:ConditionsForMAG}
\end{eqnarray}
where the symmetric mapping 
$Q_L : \{0,1\}^L \times \{0,1\}^L \rightarrow [0,1]$ was introduced earlier at (\ref{eq:Q_L}).
Thus, the probabilistic characteristics of $\mathbb{M}(n,L)$ are 
completely determined by the matrix $\mathcal{Q}$ and by the pmf $\myvec{\mu}$. 
These building blocks are assumed given and held {\em fixed} during the discussion -- They will not
be explicitly displayed in the notation.

Throughout we write
\begin{equation}
\Gamma(a) =  \bE{ q(a, A) } ,
\quad a=0,1
\label{eq:Gamma}
\end{equation}
with results all given under the compact condition $\Gamma(0) < \Gamma (1)$.
When $\Gamma(1) < \Gamma(0)$, 
the results can be obtained {\em mutatis mutandis}
by exchanging the roles of the attributes  $0$ and $1$, 
i.e.,  the roles of $\mu(0)$ (resp. $\Gamma(0)$) and $\mu(1)$
(resp. $\Gamma(1)$) need to be  interchanged in various statements.
Details are left to the interested reader.

\section{The main results}
\label{sec:Results}

Fix $n=2,3, \ldots$ and $L=1,2, \ldots $.
For each $u=1, \ldots , n$, node $u$ is {\em isolated} in $\mathbb{M}(n;L)$ if
there is no other node (in $\{1, \ldots , n \}$) distinct from $u$ which is $L$-adjacent to  node $u$.
The $\{ 0,1\}$-valued rv $\xi_{n,L}(u)$ given by
\begin{equation}
\xi_{n,L}(u) \equiv  \prod_{w=1, \ w \neq u}^n \left ( 1 - \chi_L(u,w) \right )
\label{eq:IsolatedNodeIndicator}
\end{equation}
encodes the fact that node $u$ is isolated in $\mathbb{M}(n;L)$.

We are interested in establishing a zero-one law for  the absence of isolated nodes in MAG models when
the number $n$ of nodes and the number $L$ of nodal attributes grow unboundedly large, the latter quantity
scaling with the former.
The following terminology, used repeatedly in what follows, should help simplify the presentation:
A {\em scaling} (for the number of attributes) is any mapping ${\bf L}: \mathbb{N}_0 \rightarrow \mathbb{N}_0: n \rightarrow L_n$.
With $\rho >0$, the scaling ${\bf L}: \mathbb{N}_0 \rightarrow \mathbb{N}_0$ is said to be $\rho$-{\em admissible}
if 
\begin{equation}
L_n \sim \rho \ln  n,
\label{eq:Rho-Admissible}
\end{equation}
in which case it holds that
\begin{equation}
L_n = \rho_n \ln  n,
\quad n=1,2, \ldots 
\label{eq:Rho-AdmissibleEquivalent}
\end{equation}
for some sequence  $\myvec{\rho}: \mathbb{N}_0 \rightarrow \mathbb{R}_+: n \rightarrow \rho_n$ 
such that $\lim_{n \rightarrow \infty} \rho_n = \rho$. The sequence $\myvec{\rho}: \mathbb{N}_0 \rightarrow \mathbb{R}_+$
defined by (\ref{eq:Rho-AdmissibleEquivalent}) is uniquely determined by the $\rho$-scaling ${\bf L}: \mathbb{N}_0 \rightarrow \mathbb{N}_0$,
and is said to be {\em associated} with it.

Interest in admissible scalings is discussed in \cite{KimLeskovec}.
The definition of admissibility given by Kim and Leskovec \cite{KimLeskovec} uses logarithms in base two;
results given here are easily reconciled with the ones in \cite{KimLeskovec}
through the well-known fact that $\ln x = \ln 2 \cdot \log_2 x$ with $\log_2 x $ denoting the logarithmof $x$ in base $2$ for $x > 0$.
In particular, a $\rho$-admissible scaling as defined here at (\ref{eq:Rho-AdmissibleEquivalent}) is a $\rho \ln 2$-scaling in the sense of Kim and Leskovec.

The zero-one law for the absence of isolated nodes takes a different form depending on the sign of $1 + \rho \ln  \mu(0)$.
The boundary case $1 + \rho \ln  \mu(0)=0$ will not be considered in what follows.

\subsection{The case $1 + \rho \ln  \mu(0) > 0$}

The result given next contains the zero-one law under the condition $1 + \rho \ln  \mu(0) > 0$, and is established in Section \ref{sec:ProofThmPart1}.

\begin{theorem}
{\sl
Assume $\Gamma(0) < \Gamma (1)$.
With $\rho > 0$, we further assume that
\begin{equation}
1 + \rho \ln  \mu(0) > 0.
\label{eq:BasicCondition+Part1}
\end{equation}
Then, for any $\rho$-admissble scaling
${\bf L}: \mathbb{N}_0 \rightarrow \mathbb{N}_0$, we have the zero-one law
\begin{equation}
\lim_{n \rightarrow \infty}
\bP{
\begin{array}{c}
\mbox{ $\mathbb{M}(n;L_n)$ contains}  \\
\mbox{ no isolated nodes} \\
\end{array}
}
=
\left \{
\begin{array}{ll}
0 & \mbox{if $1 + \rho \ln  \Gamma(0) < 0$ } \\
   & \\
1 & \mbox{if $1 + \rho \ln  \Gamma(0) > 0$. } \\
\end{array}
\right .
\label{eq:Zero-OneLaw+Part1}
\end{equation}
}
\label{thm:Zero-OneLaw+Part1}
\end{theorem}

\subsection{The case $1 + \rho \ln  \mu(0) < 0$}

Theorem \ref{thm:Zero-OneLaw+Part1} takes a very different form when (\ref{eq:BasicCondition+Part1}) does not hold. 
To state the results, we introduce the quantity
\begin{equation}
G(\nu,\mu) 
=
\left ( \frac{\mu}{\nu} \right )^\nu 
\left ( \frac{1-\mu}{1-\nu} \right )^{1-\nu},
\quad 0 < \nu, \mu < 1 .
\label{eq:DefnG}
\end{equation}
For each $\mu$ in $(0,1)$ the mapping $(0,1) \rightarrow \mathbb{R}_+: \nu \rightarrow G(\nu,\mu) $ is
well defined and continuous. 
By continuity we can extend it into  into a continuous mapping defined on the {\em closed} interval $[0,1]$ so that
$G(0,\mu)  = \lim_{ \nu \downarrow 0 } G(\nu,\mu) = 1 - \mu$ and $G(1,\mu)  = \lim_{ \nu \uparrow 1 } G(\nu,\mu) = \mu$.
This corresponds to using the convention $0^0 =1$ in the expression  (\ref{eq:DefnG}).
In a similar way, for each $\mu$ in $(0,1)$ the mapping $(0,1) \rightarrow \mathbb{R}: \nu \rightarrow \ln  G(\nu,\mu) $ is
well defined and continuous with
\begin{equation}
\ln  G(\nu,\mu) 
=
- \nu \ln   \left ( \frac{\nu}{\mu} \right ) -  (1-\nu) \ln  \left ( \frac{1-\nu}{1-\mu} \right ),
 \quad 0 < \nu < 1 .
 \label{eq:DefnLogG}
\end{equation}
We can also extend this second mapping into a continuous mapping defined on the closed interval $[0,1]$ with
$\ln  G(0,\mu)  = \lim_{ \nu \downarrow 0 } \ln  G(\nu,\mu) = \ln  ( 1 - \mu )$
and
$\ln  G(1,\mu)  = \lim_{ \nu \uparrow 1 } \ln  G(\nu,\mu) = \ln  \mu$.
This is consistent with applying the usual convention $0 \ln  0 = 0$ in the expression  (\ref{eq:DefnLogG}).
Elementary calculus shows that the mapping $[0,1] \rightarrow \mathbb{R}: \nu \rightarrow \ln  G(\nu,\mu) $  is concave,
and that its maximum is achieved at $\nu=\mu$ with $\ln  G(\mu,\mu) = 0$.
Thus, the mapping $[0,1] \rightarrow \mathbb{R}: \nu \rightarrow \ln  G(\nu,\mu) $ increases on $(0,\mu)$, 
reaches its maximum at $\nu=\mu$ and then decreases on $(\mu, 1)$.

With these preliminaries in place, for each $\mu$ in $(0,1)$ and $\rho > 0$, consider the non-linear equation
\begin{equation}
1 + \rho \ln  G(\nu, \mu) = 0,
\quad \nu \in [0,1].
\label{eq:EqnForNU_A}
\end{equation}
If the condition $1 + \rho \ln  (1-\mu) < 0$
holds, then the equation (\ref{eq:EqnForNU_A}) has a {\em non}-empty set of solutions. 
More precisely,  there always exists a root, denoted $\nu_\star (\rho)$, in the interval $(0,\mu)$ since
$1 + \rho \ln  G(0, \mu) = 1 + \rho \ln  (1-\mu) < 0$ while $1 + \rho \ln  G(\mu, \mu) = 1 $.
Only when
\[
1 + \rho \ln  G(1, \mu) = 1 + \rho \ln  \mu \leq 0,
\]
does there exist a second root located in the interval $(\mu, 1]$.
In what follows $\mu(1)$ plays the role of $\mu$.

\begin{theorem}
{\sl
Assume $\Gamma(0) < \Gamma (1)$.
With $\rho > 0$, we further assume that
\begin{equation}
1 + \rho \ln  \mu(0) < 0.
\label{eq:BasicCondition+Part2}
\end{equation}
Then, for any $\rho$-admissible scaling
${\bf L}: \mathbb{N}_0 \rightarrow \mathbb{N}_0$, we have the zero-one law
\begin{equation}
\lim_{n \rightarrow \infty}
\bP{
\begin{array}{c}
\mbox{ $\mathbb{M}(n;L_n)$ contains}  \\
\mbox{ no isolated nodes} \\
\end{array}
}
=
\left \{
\begin{array}{ll}
0 & \mbox{if $1 + \rho \ln  \left ( \Gamma(1)^{\nu_\star (\rho) } \Gamma(0)^{1-\nu_\star (\rho)} \right ) < 0$ } \\
   & \\
1 & \mbox{if $1 + \rho \ln  \left ( \Gamma(1)^{\nu_\star (\rho)} \Gamma(0)^{1-\nu_\star (\rho) } \right ) > 0$} \\
\end{array}
\right .
\label{eq:Zero-OneLaw+Part2}
\end{equation}
where $\nu_\star(\rho) $ is the unique solution in the interval $(0, \mu(1))$ to the equation
\begin{equation}
1 + \rho \ln  G(\nu, \mu(1)) = 0,
\quad \nu \in [0,1].
\label{eq:EqnForNU}
\end{equation}
}
\label{thm:Zero-OneLaw+Part2}
\end{theorem}

Theorem \ref{thm:Zero-OneLaw+Part2} is established in Section \ref{sec:CompleteProofThmPart2}
with the help of auxiliary results discussed in Section
\ref{sec:Proof+COMPLEMENT_Infinity-OneLaw+FirstMoment+Part2} and Section \ref{sec:Proof+Infinity-OneLaw+FirstMoment+Part2}.

\subsection{On the conditions at (\ref{eq:Zero-OneLaw+Part2})}

For future reference, in order to avoid repetitions,
we discuss the constraints on the sign of 
$1 + \rho \ln  \left ( \Gamma(1)^{\nu_\star (\rho) } \Gamma(0)^{1-\nu_\star (\rho)} \right )$
which appear in the statement of  Theorem \ref{thm:Zero-OneLaw+Part2}. 
As we will discover shortly in subsequent sections, forthcoming arguments will require
the existence of a value $\nu$ either in the range $(0,\nu_\star(\rho))$ such that 
\begin{equation}
1 + \rho \ln  \left ( \Gamma(1)^{\nu } \Gamma(0)^{1-\nu} \right ) < 0,
\label{eq:ConditionForNU-}
\end{equation}
or in the range $(\nu_\star(\rho), \mu(1))$ such that 
\begin{equation}
1 + \rho \ln  \left ( \Gamma(1)^{\nu} \Gamma(0)^{1-\nu} \right ) > 0.
\label{eq:ConditionForNU+}
\end{equation}

As we now argue, the existence of  a value $\nu$ in the requisite intervals is indeed guaranteed by
the conditions
\begin{equation}
1 + \rho \ln  \left ( \Gamma(1)^{\nu_\star (\rho) } \Gamma(0)^{1-\nu_\star (\rho)} \right ) < 0
\label{eq:ConditionForNUstar-}
\end{equation}
and
\begin{equation}
1 + \rho \ln  \left ( \Gamma(1)^{\nu_\star (\rho) } \Gamma(0)^{1-\nu_\star (\rho)} \right ) > 0,
\label{eq:ConditionForNUstar+}
\end{equation}
respectively: The elementary fact
\[
1 + \rho \ln  \left ( \Gamma(1)^{\nu} \Gamma(0)^{1-\nu} \right )
=
1 + \rho 
\left ( \nu \ln  \Gamma(1)  +  (1-\nu) \ln  \Gamma(0) \right ),
\quad \nu \in [0,1],
\]
shows that the mapping $\nu \rightarrow 1 + \rho \ln  \left ( \Gamma(1)^{\nu} \Gamma(0)^{1-\nu} \right )$
is {\em affine} (thus continuous)  on $[0,1]$ and strictly increasing (since $\Gamma(0) < \Gamma(1)$) with intercepts  at $\nu=0$ and $\nu=1$ given by
$1 + \rho \ln  \Gamma(0) $ and $1 + \rho \ln  \Gamma(1)$, respectively.
This elementary observation has the following implications:
If (\ref{eq:ConditionForNUstar-}) holds, then by continuity and monotonicity there exists a non-trivial interval
$I_-(\rho) = (\alpha_-(\rho), \beta_-(\rho))$ contained in $(0, \mu(1))$ with the following properties:
The interval $I_-(\rho)$ contains $\nu_\star(\rho)$ and (\ref{eq:ConditionForNU-}) holds on it.
On the other hand,  if (\ref{eq:ConditionForNUstar+}) holds, then again by continuity and monotonicity there now
exists a non-trivial interval
$I_+(\rho) = (\alpha_+(\rho), \beta_+(\rho))$ contained in $(0, \mu(1))$ such that $\nu_\star(\rho)$ belongs to $I_+(\rho)$
and (\ref{eq:ConditionForNU+}) holds on it.

Finally, we close by noting that
Kim and Leskovec couch their analysis in terms of the counts
\[
\sum_{u \in V_n} \1{ L- S_L(u) = j } ,
\quad j=0, \ldots , L
\]
while here we have used instead the counts
\[
\sum_{u \in V_n} \1{ S_L(u) = j } ,
\quad j=0, \ldots , L.
\]
In other words, Kim and Leskovec count the NO answers while we count the YES answers.
This is why the parameters $\mu(0)$ and $\mu(1)$ need to be exchanged to go from the conditions appearing in their paper to the
ones appearing here. However, Leskovec and Kim do impose {\em additional} conditions 
on the entries of the symmetric matrix $\mathcal{Q}$, namely that $q(1,1) < q(0,1) = q(1,0)  < q(0,0)$
(so that $\Gamma (1) < \Gamma(0)$ with their convention). 
Here we ask only for $\Gamma (0) < \Gamma(1)$ (with our conventions) with no additional conditions.

\section{A roadmap to the proofs}
\label{sec:Roadmap}

\subsection{Counting isolated nodes}
\label{subsec:CountingIsolatedNodes}

Fix $n=2,3, \ldots$ and $L=1,2, \ldots $.
To count  the number of isolated nodes in $\mathbb{M}(n;L)$ we introduce the rv $I_n(L)$ given by
\begin{equation}
I_n(L)
= \sum_{u=1}^n \xi_{n,L}(u) .
\label{eq:IsolatedNodesCount}
\end{equation}
Interest in these count variables stems from the observation that
$\mathbb{M}(n;L)$ contains no isolated nodes if and only if $I_n(L) = 0$, and that te key relation
\begin{equation}
\bP{~ \mbox{ $\mathbb{M}(n;L)$ contains no isolated nodes}~ }
=
\bP{ I_n(L) = 0 }
\label{eq:KeyRelation}
\end{equation}
holds.
This fact will be used to establish Theorems \ref{thm:Zero-OneLaw+Part1} and  \ref{thm:Zero-OneLaw+Part2}
by leveraging easy bounds on the probability $\bP{ I_n(L) = 0 }$ in terms of 
the first and second moments of the rv $I_n(L)$ (as discussed next in Section \ref{subsec:MethodFirstSecondMoment}).

However, some of the forthcoming arguments will require a finer accounting which we now introduce.
Recall that for each node $u=1, \ldots , n$, the number of attributes exhibited by node $u$ amongst the first $L$ attributes
is captured by the rv $S_L(u)$ introduced at (\ref{eq:CountAttributes}).
For each $\ell =0,1, \ldots , L$, the $\{0,1\}$-valued rv $\xi^{(\ell)}_{n,L} (u) $ given by
\begin{equation}
\xi^{(\ell)}_{n,L} (u) =  \xi_{n,L} (u) \cdot \1{ S_L(u) = \ell }.
\label{eq:IsolatedNodeIndicatorWithEllAttributes}
\end{equation}
indicates whether node $u$ is isolated in $\mathbb{M}(n;L)$ while $\ell$ attributes
are present amongst its first $L$ attributes.

The total number of isolated nodes in $\mathbb{M}(n;L)$ which have $\ell$ attributes amongst the first $L$ attributes
is then given by
\begin{equation}
I^{(\ell)} _n (L) 
= \sum_{u=1}^n \xi^{(\ell)}_{n,L} (u) 
=  \sum_{u=1}^n \xi_{n,L} (u) \1{ S_L(u) = \ell }.
\label{eq:IsolatedNodesCountWithEllAttributes}
\end{equation}
Simple accounting readily yields the relations
\begin{equation}
\xi_{n,L} (u) =   \sum_{\ell=0}^L  \xi^{(\ell)}_{n,L} (u)
\label{eq:Accounting1}
\end{equation}
and
\begin{equation}
I_n(L) = \sum_{\ell=0}^L I^{(\ell)} _n (L),
\label{eq:Accounting2}
\end{equation}
the last one yielding the elementary bounds
\begin{equation}
I^{(\ell)} _n (L)  \leq I_n(L),
\quad \ell=0,1, \ldots , L.
\label{eq:EasyBounds}
\end{equation}

\subsection{The method of first and second moments}
\label{subsec:MethodFirstSecondMoment}

The basic strategy for proving Theorems \ref{thm:Zero-OneLaw+Part1} and  \ref{thm:Zero-OneLaw+Part2}
relies on the method of first and second moments applied to the number (\ref{eq:IsolatedNodesCount})
of isolated nodes and to the related count variables (\ref{eq:IsolatedNodesCountWithEllAttributes}).
In this section we provide the main ingredients of this approach as we will need it in its various applications.

Let $\{ Z_n , \ n=1,2, \ldots \}$ denote a collection of $\mathbb{N}$-valued rvs such that
$\bE{ Z_n^2} < \infty$ for each $n=1,2, \ldots$.
The method of first moment 
\cite[Eqn (3.10), p. 55]{JansonLuczakRucinski} relies on the well-known bound
\begin{equation}
1 - \bE{ Z_n } \leq \bP{  Z_n = 0 }
\label{eq:BoundViaFirstMomentZ}
\end{equation}
while the method of second moment 
\cite[Remark 3.1, p. 55]{JansonLuczakRucinski} has its starting point in the inequality
\begin{equation}
\bP{  Z_n = 0 } 
\leq 1 - \frac{ \left ( \bE{ Z_n } \right )^2}{
\bE{ Z_n^2} }. 
\label{eq:BoundViaSecondMomentZ}
\end{equation}

Letting $n$ go to infinity in the resulting inequalities, we conclude from
(\ref{eq:BoundViaFirstMomentZ}) that
\begin{equation}
\lim_{n \rightarrow \infty} 
\bP{ Z_n = 0 }  = 1
\label{eq:OneLawZ}
\end{equation}
if
\begin{equation}
\lim_{n \rightarrow \infty} \bE{ Z_n } = 0,
\label{eq:FirstMomentConditionZ}
\end{equation}
while the bound (\ref{eq:BoundViaSecondMomentZ}) implies
\begin{equation}
\lim_{n \rightarrow \infty}  \bP{ Z_n = 0 }  = 0
\label{eq:ZeroLawZ}
\end{equation}
whenever
\begin{equation}
 \limsup_{n \rightarrow \infty}   
\frac{ \bE{ Z_n^2 } }
        { \left ( \bE{ Z_n } \right )^2 }
\leq 1.
\label{eq:SecondMomentConditionZ}
\end{equation}

Here we use this strategy when the rvs $\{ Z_n , \ n=1,2, \ldots \}$ are count variables 
with the following structure: For  each $n=1,2, \ldots $,
the rv $Z_n$ is of the form
\[
Z_n = \sum_{u=1}^n \zeta_{n}(u)
\]
where the rvs $\zeta_{n}(1), \ldots , \zeta_{n}(n)$ are $\{0,1\}$-valued rvs. 
If in addition, the rvs $\zeta_{n}(1), \ldots , \zeta_{n}(n)$ are {\em exchangeable}
(as they will be here), then we easily arrive at the expressions
\begin{eqnarray}
\bE{ Z_n }
=
\bE{ \sum_{u=1}^n \zeta_{n} (u) }
= n \bE{ \zeta_{n}(1) }
\label{eq:FirstMomentA}
\end{eqnarray}
and
\begin{eqnarray}
\bE{ Z_n^2 }
=
\bE{ \left ( \sum_{u=1}  \zeta_{n}(u) \right )^2 }
=
n \bE{ \zeta_{n,1} }  + n(n-1) \bE{ \zeta_{n}(1) \cdot  \zeta_{n}(2) }
\end{eqnarray}
by virtue of the binary nature of the rvs involved, whence
\begin{eqnarray}
\frac{ \bE{ Z_n ^2 } }
        { \left ( \bE{ Z_n } \right )^2 }
        =
        \frac{1}{  \bE{ Z_n } } + \frac{n-1}{n} \cdot
        \frac{ \bE{ \zeta_{n} (1) \cdot \zeta_{n}(2)  }  }{ \left ( \bE{\zeta_{n}(1) } \right )^2 }.
\end{eqnarray}

It is now plain that (\ref{eq:SecondMomentConditionZ}) 
can be achieved if we show that
\begin{equation}
\lim_{n \rightarrow \infty} 
\bE{ Z_n } = \infty,
\label{eq:FirstMomentConditionAlsoZ}
\end{equation}
and
\begin{equation}
\limsup_{n \rightarrow \infty}   
\frac{ \bE{ \zeta_{n}(1) \cdot  \zeta_{n}(2)  }  }{ \left ( \bE{\zeta_{n}(1)  } \right )^2 }
\leq 1.
\label{eq:SecondMomentConditionAlsoZ}
\end{equation}

For the problem at hand,  we shall proceed as follows:
With a $\rho$-scaling ${\bf L}: \mathbb{N}_0 \rightarrow \mathbb{N}_0$ for some $\rho > 0$,
we seek to establish  the desired zero-one laws through the convergence  $\lim_{n \rightarrow \infty} \bP{ I_n(L_n) = 0 } = 0$
and $\lim_{n \rightarrow \infty} \bP{ I_n(L_n) = 0 } = 1$.
In principle this could be achieved by applying the method of first and second moments to the rvs $\{ Z_n, \ n=1,2, \ldots \}$ given by
\begin{equation}
Z_n = I_n(L_n),
\quad n=1,2, \ldots 
\label{eq:Z=I}
\end{equation}
However, while this approach will work quite easily for the one-law, we will encounter some difficulty in
applying the method of second moment to the rvs (\ref{eq:Z=I}) and a somewhat indirect approach 
(based on (\ref{eq:EasyBounds})) will be adopted.

\section{Evaluating the first two moments}
\label{sec:EvaluatingMoments}

\subsection{Evaluating the first moments}
\label{subsec:FirstMoments}

We begin with an easy calculation of the first moments.

\begin{lemma}
{\sl Consider arbitrary $n=2,3, \ldots $ and $L=1,2, \ldots$.
For each $u=1, \ldots , n$, with $S_L(u)$ given by (\ref{eq:CountAttributes}), it holds that
\begin{equation}
\bE{ \xi^{(\ell)}_{n,L} (u) }
=  \left ( 1 - \Gamma(1)^\ell \Gamma(0)^{L-\ell} \right )^{n-1} \cdot  \bP{ S_L(u)  = \ell },
\quad \ell = 0,1, \ldots , L
\label{eq:EvalFirstMomentWithEll}
\end{equation}
and
\begin{equation}
\bE{ \xi_{n,L} (u) }
=  \bE{ \left ( 1 - \Gamma(1)^{S_L(u)} \Gamma(0)^{L-S_L(u)} \right )^{n-1} }.
\label{eq:EvalFirstMoment}
\end{equation}
}
\label{lem:EvalFirstMoment}
\end{lemma}

Recall that the rvs $\{ A, A_\ell, \ \ell =1, 2, \ldots \}$ are i.i.d. $\{0,1\}$-valued rvs with 
pmf  $\myvec{\mu}$, and corresponding sequence of partial sums $\{ S_L, \ L =1,2, \ldots \}$
given by (\ref{eq:CountAttributesGeneric}).
Under the enforced Assumptions (i)-(iii) it is plain that for each $L=1,2, \ldots $,
the rvs $S_L(1),  S_L(2) , \ldots,  S_L(n)$ are i.i.d., each distributed according to the rv $S_L$.
The two relations
\begin{equation}
\bE{ I^{(\ell)}_{n} (L) }
=  n
\left ( 1 -  \Gamma(1)^\ell \Gamma(0)^{L-\ell} \right )^{n-1}  \cdot  \bP{ S_L  = \ell },
\quad \ell = 0,1, \ldots , L
\label{eq:IwithEll}
\end{equation}
and
\begin{equation}
\bE{ I_{n} (L) }
=  n  \bE{ \left ( 1 - \Gamma(1)^{S_L} \Gamma(0)^{L-S_L} \right )^{n-1}  }
\label{eq:I}
\end{equation}
are now immediate consequences of the relations
(\ref{eq:IsolatedNodesCountWithEllAttributes}) and  (\ref{eq:Accounting2}), respectively.

In what follows, for each $L=1,2, \ldots$,  we shall have use for the moments
\begin{equation}
Q^\star_L (\myvec{a}_L )
=
\bE{ Q_L \left ( \myvec{a}_L,    \myvec{A}_L  \right ) },
\quad \myvec{a}_L \in \{0,1\}^L.
\label{eq:Q_StarL}
\end{equation}
Note that
\begin{eqnarray}
Q^\star_L (\myvec{a}_L )
= \bE{ \prod_{\ell=1}^L q(a_\ell , A_\ell ) } 
= \bE{ q(1, A) }^{\sum_{\ell=1}^L a_\ell} \cdot  \bE{ q( 0, A) }^{\sum_{\ell=1}^L (1-a_\ell)}
\label{eq:Q_StarLagain}
\end{eqnarray}
as we use the fact that the $\{0,1\}$-valued rv $A$ is a generic representative  of the i.i.d. rvs $A_1, \ldots , A_L$. 
In particular it follows that
\begin{equation}
Q^\star_L (\myvec{A}_L(u) ) = \Gamma(1)^{S_L(u)} \Gamma(0)^{L-S_L(u)},
\quad u=1, \ldots , n.
\label{eq:BasicIdentity}
\end{equation}

\myproof
It suffices to show that (\ref{eq:EvalFirstMomentWithEll}) holds
since (\ref{eq:EvalFirstMoment}) follows as an easy consequence 
of the expression (\ref{eq:Accounting1}).
Pick positive $n=2,3, \ldots $ and $L=1,2, \ldots $, 
and consider node $u=1, \ldots , n$.
For each $\ell  =0,1, \ldots , L$, 
with the relation (\ref{eq:IsolatedNodeIndicatorWithEllAttributes})
holding, a standard preconditioning argument yields
\begin{equation}
\bE{  \xi^{(\ell)}_{n,L} (u) }
=
\bE{
\1{ S_L(u) = \ell }    \cdot\bE{  \xi_{n,L}(u) \Bigl |  \myvec{A}_L (u) }
}
\label{eq:FirstMoment=A}
\end{equation}
as we note that the rv $S_L(u)$ is determined by the attribute vector $\myvec{A}_L (u)$.

With (\ref{eq:IsolatedNodeIndicator}) as a point of departure, we have
\begin{eqnarray}
\xi_{n,L}(u)
=
\prod_{w=1, \ w \neq u}^n \left ( 1 - \chi_{L}(u,w) \right )
=
\prod_{w=1, \ w \neq u}^n
\1{ U(u,w) > Q_L ( \myvec{A}_L(u), \myvec{A}_L(w) ) }  .
\nonumber
\end{eqnarray}
Under the enforced independence assumptions, we readily conclude to
\begin{eqnarray}
\bE{  \xi_{n,L}(u)  \Bigl |  \myvec{A}_L (1), \ldots ,  \myvec{A}_L (n) }
=
\prod_{w=1, \ w \neq u }^n \left ( 1 - Q_L ( \myvec{A}_L (u), \myvec{A}_L (w) ) \right ) .
\nonumber 
\end{eqnarray}
The smoothing property of conditional expectations readily gives
\begin{eqnarray}
\bE{  \xi_{n,L}(u)  \Bigl |  \myvec{A}_L (u) }
&=&
\bE{ \bE{  \xi_{n,L}(u)  \Bigl |  \myvec{A}_L (1), \ldots ,  \myvec{A}_L (n) } \Bigl |  \myvec{A}_L (u)  }
\nonumber \\
&=&
\bE{ \prod_{w=1, \ w \neq u }^n \left ( 1 - Q_L ( \myvec{A}_L (u), \myvec{A}_L (w) ) \right )
\Biggl |  \myvec{A}_L (u) }
\nonumber \\
&=&
\bE{ \prod_{w=1, \ w \neq u }^n \left ( 1 - Q_L ( \myvec{a}_L, \myvec{A}_L (w) ) \right ) }_{ \myvec{a}_L = \myvec{A}_L (u) }
\nonumber \\
&=&
\left (
1 - Q^{\star}_L( \myvec{A_L}(u) )
\right )^{n-1}
\nonumber 
\end{eqnarray}
where the last two steps made use of the fact that the rvs $\myvec{A}_L (1), \ldots ,  \myvec{A}_L (n)$ are i.i.d. rvs.
Using (\ref{eq:FirstMoment=A}) we obtain
\begin{eqnarray}
\bE{  \xi^{(\ell)}_{n,L} (u) }
&=&
\bE{
\1{ S_L(u) = \ell } \cdot \left ( 1 - Q^{\star}_L( \myvec{A_L}(u) ) \right )^{n-1} }
\nonumber \\
&=&
\bE{
\1{ S_L(u) = \ell } \cdot \left ( 1 - \Gamma(1)^{S_L(u)} \Gamma(0)^{L-S_L(u)}  \right )^{n-1} 
}
\nonumber 
\end{eqnarray}
by virtue of (\ref{eq:BasicIdentity}), and the desired conclusion (\ref{eq:EvalFirstMomentWithEll})  follows in a straightforward manner.
\myendpf

\subsection{Evaluating the second moments}
\label{subsec:SecondMoments}

The expressions for the second order quantities are much more involved as the next intermediary result already shows. 

\begin{lemma}
{\sl Consider arbitrary $n=2,3, \ldots $ and $L=1,2, \ldots$.
For distinct $u,v=1, \ldots , n$, it holds that
\begin{eqnarray}
\lefteqn{
\bE{  \xi_{n,L}(u)  \xi_{n,L}(v) \Bigl |  \myvec{A}_L (u),  \myvec{A}_L (v) }
} & & 
\label{eq:EvalSecondMomentConditional} \\
&=&
\left ( 1 - Q_L ( \myvec{A}_L (u), \myvec{A}_L (v) ) \right )
\cdot 
\left ( 1 - \widetilde Q_L ( \myvec{A}_L (u), \myvec{A}_L (v) \right )^{n-2}
\nonumber
\end{eqnarray}
where for  arbitrary $\myvec{a_L}$ and $\myvec{b}_L$ in $ \{ 0,1 \}^L $,  we have set
\begin{equation}
\widetilde Q_L ( \myvec{a_L}, \myvec{b}_L )
=
Q^{\star}_L( \myvec{a}_L  ) + Q^{\star}_L( \myvec{b}_L)  -  Q^{\star\star}_L(   \myvec{a_L}, \myvec{b}_L  )
\label{eq:WideTildeQ}
\end{equation}
with
\begin{equation}
Q^{\star\star}_L( \myvec{a_L}, \myvec{b}_L )
= \bE{ Q_L ( \myvec{a}_L , \myvec{A}_L )  Q_L ( \myvec{b}_L , \myvec{A}_L )  }.
\label{eq:Q^StarStar}
\end{equation}
}
\label{lem:EvalSecondMoment}
\end{lemma}

The proof of this result can be found in Appendix \ref{app:ProofLemmaEvalSecondMoment}.
In principle,  it is now possible to evaluate the expressions
\[
\bE{  \xi^{(k)}_{n,L}(u) \xi^{(\ell)}_{n,L}(v) },
\quad k,\ell =0, \ldots , L
\]
for distinct $u,v=1, \ldots , n$.
Indeed, for $k, \ell  =0,1, \ldots , L$, not necessarily distinct, 
the relation (\ref{eq:IsolatedNodeIndicatorWithEllAttributes}) yields
\begin{eqnarray}
\xi^{(k)}_{n,L}(u) \xi^{(\ell)}_{n,L}(v)
=
\1{ S_L(u) = k }   \1{ S_L(v) = \ell }   \cdot \xi_{n,L} (u)   \xi_{n,L} (v) 
\label{eq:ProductWithKandL}
\end{eqnarray}
and an easy preconditioning argument leads to
\begin{eqnarray}
\lefteqn{
\bE{  \xi^{(k)}_{n,L} (u) \cdot  \xi^{(\ell)}_{n,L} (v) }
} & &
\nonumber \\
&=&
\bE{
\1{ S_L(u) = k }   \1{ S_L(v) = \ell }   
\cdot
\bE{  \xi_{n,L}(u)  \xi_{n,L}(v) \Bigl |  \myvec{A}_L (u), \myvec{A}_L (v) }
}
\label{eq:ExpectedCrossProduct}
\end{eqnarray}
because the rvs $S_L(u)$ and $S_L(v)$ are determined 
by the attribute vectors $\myvec{A}_L (u)$ and $\myvec{A}_L (v)$, respectively.
Using (\ref{eq:Accounting1}) we also readily obtain
\[
\bE{  \xi_{n,L}(u) \xi_{n,L}(v) } 
= \sum_{k=0}^L \sum_{\ell=0}^L \bE{  \xi^{(k)}_{n,u}(L) \xi^{(\ell)}_{n,L}(v) }.
\]

With arbitrary $\myvec{a_L}$ and $\myvec{b}_L$ in $ \{ 0,1 \}^L $, we note from (\ref{eq:Q^StarStar}) that
\begin{eqnarray}
\lefteqn{
Q^{\star\star}_L( \myvec{a_L}, \myvec{b}_L )
} & &
\nonumber \\
&=&
 \bE{ Q_L ( \myvec{a}_L , \myvec{A}_L )  Q_L ( \myvec{b}_L , \myvec{A}_L )  }
 \nonumber \\
 &=&
 \bE{ \prod_{\ell=1}^L q ( a_\ell , A_\ell )  q ( b_\ell , A_\ell )    }
 \nonumber \\
 &=&
 \prod_{\ell=1}^L  \bE{  q ( a_\ell , A_\ell )  q ( b_\ell , A_\ell )    }
 \nonumber \\
 &=&
 \prod_{\ell=1}^L  \bE{  q ( 1 , A  )^2  }^{a_\ell b_\ell}  \bE{  q ( 1 , A  )  q ( 0 , A )    }^{ a_\ell (1-b_\ell)+ b_\ell (1-a_\ell)}  
 \bE{  q ( 0 , A  ) ^2   }^{ (1-a_\ell )(1- b_\ell )} 
 \nonumber \\
 &=&
 \bE{  q ( 1 , A  )^2  }^{\sum_{\ell=1}^L a_\ell b_\ell}  \bE{  q ( 1 , A  )  q ( 0 , A )    }^{ \sum_{\ell=1}^L  a_\ell (1-b_\ell)+ b_\ell (1-a_\ell)}  
 \bE{  q ( 0 , A  ) ^2   }^{ \sum_{\ell=1}^L (1-a_\ell )(1- b_\ell )} 
 \nonumber
\end{eqnarray}
by arguments similar to the ones used for reaching  the expression (\ref{eq:Q_StarLagain}).
Here lies the rub: The quantities $Q^{\star}_L( \myvec{A_L}(u) )$ and $Q^{\star}_L( \myvec{A_L}(v) )$ depend on
$\myvec{A_L}(u)$ and $\myvec{A_L}(v)$ {\em only} through the sums $S_L(u)$ and $S_L(v)$, respectively, 
On the other hand, $Q^{\star\star}_L( \myvec{A_L}(u), \myvec{A}_L (v) )$ does {\em not} depend on
$\myvec{A_L}(u)$ and $\myvec{A_L}(v)$ only through the sums $S_L(u)$ and $S_L(v)$, 
but instead through the three sums
$\sum_{\ell=1}^L A_\ell(u) A_\ell (v)$, $\sum_{\ell=1}^L \left ( A_\ell(u) \left ( 1 - A_\ell (v) \right ) + A_\ell(v) \left ( 1 - A_\ell (u) \right ) \right )$
and
$\sum_{\ell=1}^L \left ( 1 - A_\ell(u) \right ) \left ( 1 - A_\ell (v) \right )$.

Fortunately, the {\em exact} expression (\ref{eq:EvalSecondMomentConditional}) will not be needed 
as only the following crude bounds will suffice:
For $k,\ell =0,1, \ldots , L$, not necessarily distinct, the expression 
(\ref{eq:ExpectedCrossProduct}) yields the bound
\begin{eqnarray}
\bE{  \xi^{(k)}_{n,L} (u) \cdot  \xi^{(\ell)}_{n,L} (v) }
&\leq &
\bP{ S_L(u) = k , S_L(v) = \ell }   
\nonumber \\
&=& \bP{ S_L(u) = k } \bP{ S_L(v) = \ell }  
\label{eq:BoundOnExpectedCrossProduct}
\end{eqnarray}
since
\[
\bE{  \xi_{n,L}(u)  \xi_{n,L}(v) \Bigl |  \myvec{A}_L (u), \myvec{A}_L (v) } \leq 1
\quad a.s.
\]

\section{Two useful technical results}
\label{sec:TwoTechnicalResults}

The next two technical lemmas will be useful in a number of places.
We present them here, with their proofs, for easy reference.
The first one relies on the following well-known fact  \cite[Prop. 3.1.1, p. 116]{EKM} : For any sequence
${\bf a}: \mathbb{N}_0 \rightarrow \mathbb{R}_+$, we have
\begin{equation}
\lim_{n \rightarrow \infty} \left ( 1 - a_n \right  )^{n} = e^{-c}
\label{eq:LimitExponentialA}
\end{equation}
for some $c$ in $[0,\infty]$ if and only if
\begin{equation}
\lim_{n \rightarrow \infty} n a_n  = c.
\label{eq:LimitExponentialB}
\end{equation} 

\begin{lemma}
{\sl Consider a $\rho$-admissible scaling ${\bf L}: \mathbb{N}_0 \rightarrow \mathbb{N}_0$ for some $\rho >0$.
For any sequence $\myvec{\nu}: \mathbb{N}_0 \rightarrow [0,1]$ such that
$\lim_{n \rightarrow \infty } \nu_n = \nu$ for some $\nu$, it holds that
\begin{eqnarray}
\lim_{n \rightarrow \infty}
\left ( 1 - \left ( \Gamma(1)^{\nu_n} \Gamma(0)^{1-\nu_n} \right )^{L_n} \right )^{n-1} 
=
\left \{
\begin{array}{ll}
0 & \mbox{if $1 + \rho \ln  \left ( \Gamma(1)^\nu \Gamma(0)^{1-\nu} \right ) > 0$} \\
       \\
 1 &  \mbox{if $1 + \rho \ln  \left ( \Gamma(1)^\nu \Gamma(0)^{1-\nu} \right )  < 0$.}  \\
\end{array}
\right .
\label{eq:TechnicalA}
\end{eqnarray}
}
\label{lem:TechnicalA}
\end{lemma}

\myproof
It follows from the equivalence 
(\ref{eq:LimitExponentialA})-(\ref{eq:LimitExponentialB})
(with $a_n =  \left ( \Gamma(1)^{\nu_n} \Gamma(0)^{1-\nu_n} \right )^{L_n}$ for all $n=1,2, \ldots $) that
the convergence
\begin{equation}
\lim_{n \rightarrow \infty}
\left ( 1 - \left ( \Gamma(1)^{\nu_n} \Gamma(0)^{1-\nu_n} \right )^{L_n}
\right )^{n-1} 
=  e^{-c}
\label{eq:LimitA1}
\end{equation}
takes place for some $c$ in $[0,\infty]$
if and only if
\begin{equation}
\lim_{n \rightarrow \infty}
(n-1) \left ( \Gamma(1)^{\nu_n} \Gamma(0)^{1-\nu_n} \right )^{L_n} = c.
\label{eq:LimitA2}
\end{equation}
For each $n=1,2, \ldots$, the $\rho$-admissibility of the scaling
${\bf L}: \mathbb{N}_0 \rightarrow \mathbb{N}_0$ yields
\begin{eqnarray}
(n-1) \left ( \Gamma(1)^{\nu_n} \Gamma(0)^{1-\nu_n} \right )^{L_n} 
&=& (n-1)  \left ( \Gamma(1)^{\nu_n} \Gamma(0)^{1-\nu_n} \right )^{\rho_n \ln  n }
\nonumber \\
&=& \frac{n-1}{n}
e^{ ( 1 + \rho_n \ln  \left ( \Gamma(1)^{\nu_n} \Gamma(0)^{1-\nu_n} \right ) ) \ln  n}
\nonumber \\
&=& \frac{n-1}{n}
n^{ 1 + \rho_n \ln  \left ( \Gamma(1)^{\nu_n} \Gamma(0)^{1-\nu_n} \right )}
\label{eq:XY}
\end{eqnarray}
where the sequence $\myvec{\rho}: \mathbb{N}_0 \rightarrow \mathbb{R}_+$ is
the unique sequence associated with the $\rho$-admissible scaling $\myvec{L}: \mathbb{N}_0 \rightarrow \mathbb{N}_0$.

The conclusion (\ref{eq:TechnicalA}) readily follows from the equivalence
of (\ref{eq:LimitA1}) and (\ref{eq:LimitA2}) once we note that
\[
\lim_{n \rightarrow \infty}
\left ( 1 + \rho_n \ln  \left ( \Gamma(1)^{\nu_n} \Gamma(0)^{1-\nu_n} \right ) \right )
= 1 + \rho \ln  \left ( \Gamma(1)^{\nu} \Gamma(0)^{1-\nu} \right ) .
\]
Indeed $1 + \rho \ln  \left ( \Gamma(1)^{\nu} \Gamma(0)^{1-\nu} \right ) < 0$
(resp. $1 + \rho \ln  \left ( \Gamma(1)^{\nu} \Gamma(0)^{1-\nu} \right ) >0$)
yields $c=0$ (resp. $c=\infty$) in (\ref{eq:LimitA2}), whence $e^{-c} = 1$ (resp. $e^{-c} = 0$)
in (\ref{eq:LimitA1}).
\myendpf

A little more can be extracted from the arguments given above:
The usual exponentiation argument and (\ref{eq:XY}) readily yield
\[
n \left ( 1 - \left ( \Gamma(1)^{\nu_n} \Gamma(0)^{1-\nu_n} \right )^{L_n} \right )^{n-1} 
\leq e^{\ln  n - \frac{n-1}{n} n^{ 1 + \rho_n \ln  \left ( \Gamma(1)^{\nu_n} \Gamma(0)^{1-\nu_n} \right )} }
\]
for $n=1,2, \ldots$. Therefore, when $1 + \rho \ln  \left ( \Gamma(1)^{\nu} \Gamma(0)^{1-\nu} \right ) > 0$, the stronger result 
\begin{equation}
\lim_{n \rightarrow \infty}
n \left ( 1 - \left ( \Gamma(1)^{\nu_n} \Gamma(0)^{1-\nu_n} \right )^{L_n} \right )^{n-1} = 0
\label{eq:XZ}
\end{equation}
also holds.

\begin{lemma}
{\sl Consider a $\rho$-admissible scaling ${\bf L}: \mathbb{N}_0 \rightarrow \mathbb{N}_0$
for some $\rho >0$.
For any sequence ${\bf C}: \mathbb{N}_0 \rightarrow (0,\infty)$ such that $\lim_{n \rightarrow \infty} C_n = C$ for
some $C>0$, it holds that
\begin{eqnarray}
\lim_{n \rightarrow \infty} nC_n^{L_n}
=
\left \{
\begin{array}{ll}
\infty  & \mbox{if $1 + \rho \ln  C > 0$} \\
       \\
0 & \mbox{if $1 + \rho \ln  C < 0$.} \\
\end{array}
\right .
\label{eq:TechnicalB}
\end{eqnarray}
}
\label{lem:TechnicalB}
\end{lemma}

\myproof
The $\rho$-admissibility of the scaling
${\bf L}: \mathbb{N}_0 \rightarrow \mathbb{N}_0$ yields
\begin{eqnarray}
n C_n^{L_n} 
= n e^{L_n \ln  C_n }
= n e^{\rho_n \ln  C_n \cdot  \ln  n }
= n^{1 + \rho_n \ln  C_n },
\quad n=2,3, \ldots
\label{eq:TechnicalB_bound}
\end{eqnarray}
where the sequence $\myvec{\rho}: \mathbb{N}_0 \rightarrow \mathbb{R}_+$ is
the unique sequence associated with the $\rho$-admissible scaling $\myvec{L}: \mathbb{N}_0 \rightarrow \mathbb{N}_0$.
Letting $n$ go to infinity readily yields the desired conclusion (\ref{eq:TechnicalB})
since $\lim_{n \rightarrow \infty} \left ( 1 + \rho_n \ln  C_n \right ) = 1 + \rho \ln  C$.
\myendpf

\section{A proof of Theorem \ref{thm:Zero-OneLaw+Part1}}
\label{sec:ProofThmPart1}

The proof of Theorem \ref{thm:Zero-OneLaw+Part1} proceeds in two steps.
The first step deals with the first moment conditions (\ref{eq:FirstMomentConditionZ})
and (\ref{eq:FirstMomentConditionAlsoZ}), and is contained in the following
\lq\lq zero-infinity" law for the first moment -- Note the analogy with Theorem \ref{thm:Zero-OneLaw+Part1}.

\begin{proposition}
{\sl
Assume $\Gamma(0) < \Gamma (1)$. With $\rho > 0$, we  further assume that (\ref{eq:BasicCondition+Part1}) holds.
For any $\rho$-admissble scaling
${\bf L}: \mathbb{N}_0 \rightarrow \mathbb{N}_0$, we have
\begin{eqnarray}
\lim_{n \rightarrow \infty} \bE{ I_n(L_n) }
=
\left \{
\begin{array}{ll}
\infty & \mbox{if $1 + \rho \ln  \Gamma(0) < 0$ } \\
         & \\
0      & \mbox{if $1 + \rho \ln  \Gamma(0) > 0$. } \\
\end{array}
\right .
\label{eq:Infinity-OneLaw+FirstMoment+Part1}
\end{eqnarray}
}
\label{prop:Infinity-OneLaw+FirstMoment+Part1}
\end{proposition}

\myproof
Fix $n=2,3, \ldots$.
Under the assumed inequality $\Gamma(0) < \Gamma(1)$, the expression (\ref{eq:I}) implies
\begin{eqnarray}
\bE{ I_n(L)}
&\leq&
n \left ( 1 - \Gamma (0) ^{L} \right )^{n-1}
\nonumber \\
&\leq& ne^{ -(n-1) \Gamma (0) ^{L}  }
\nonumber \\
&=& e^{ \ln  n - (n-1) \Gamma (0) ^{L}  },
\quad L=1,2, \ldots
\label{eq:UpperBoundFirstMoment}
\end{eqnarray}
Now, for any $\rho$-admissible scaling
${\bf L}: \mathbb{N}_0 \rightarrow \mathbb{N}_0$ we have
\begin{eqnarray}
\bE{ I_n(L_n)}
&\leq& e^{ \ln  n - (n-1) \Gamma (0) ^{L_n}  }
\label{eq:UpperBoundFirstMomentA}
\end{eqnarray}
with
\begin{eqnarray}
\ln  n - (n-1) \Gamma (0) ^{L_n}  
=
\ln  n - (n-1)  \Gamma (0) ^{\rho_n \ln  n}
=
 \ln  n - \frac{n-1}{n} n^{ 1 + \rho_n \ln  \Gamma(0) }
\label{eq:UpperBoundFirstMomentB}
\end{eqnarray}
where the sequence $\myvec{\rho}: \mathbb{N}_0 \rightarrow \mathbb{R}_+$ is
the unique sequence associated with the $\rho$-admissible scaling $\myvec{L}: \mathbb{N}_0 \rightarrow \mathbb{N}_0$.
Under the condition $1 + \rho \ln  \Gamma(0) > 0$, we have
\[
\lim_{n \rightarrow \infty} 
\left ( \ln  n - (n-1) \Gamma (0) ^{L_n}  \right ) = -\infty
\]
and the conclusion $\lim_{n \rightarrow \infty} \bE{ I_n(L_n)} = 0$ 
follows upon letting $n$ go to infinity in (\ref{eq:UpperBoundFirstMomentA}).

We now consider the case $1 + \rho \ln  \Gamma(0) < 0$: 
Fix $n=2,3, \ldots $. For each $L=1,2, \ldots $, the bound (\ref{eq:EasyBounds})
(with $\ell=0$) yields
\begin{equation}
\bE{ I^{(0)}_n(L)}
= n \left ( 1 -  \Gamma(0)^{L} \right )^{n-1}  \cdot  \bP{ S_L  = 0}
\leq \bE{ I_{n} (L) }
\nonumber 
\end{equation}
as we make use of (\ref{eq:IwithEll}) (with $\ell=0$). Recall that $\bP{ S_L  = 0} = \mu(0)^L$
since $S_L$ is a binomial rv ${\rm Bin}(L, \mu(1))$.
Now, for any $\rho$-admissible scaling ${\bf L}: \mathbb{N}_0 \rightarrow \mathbb{N}_0$ we can write
\begin{equation}
\bE{ I^{(0)}_n(L_n)}
= 
n \mu(0)^{L_n}  
\left 
( 1 - \Gamma(0)^{L_n}\right )^{n-1} \leq \bE{ I_n(L_n) } .
\label{eq:LowerBoundFirstMomentA}
\end{equation}
Let $n$ go to infinity in (\ref{eq:LowerBoundFirstMomentA}):
Lemma \ref{lem:TechnicalA} (with $\nu_n =0$ for all $n=1,2, \ldots$) gives
$\lim_{n \rightarrow \infty}  \left ( 1 - \Gamma(0)^{L_n}\right )^{n-1} = 1$ under the
condition $1 + \rho \ln  \Gamma(0) < 0$, while 
Lemma \ref{lem:TechnicalB} (with $C_n= \mu(0)$ for all $n=1,2, \ldots$) yields
$\lim_{n \rightarrow \infty} n\mu(0)^{L_n} = \infty $ under (\ref{eq:BasicCondition+Part1}).
Thus, $\lim_{n \rightarrow \infty} \bE{ I_n^{(0)}(L_n) } = \infty$, and the desired conclusion $\lim_{n \rightarrow \infty} \bE{ I_n(L_n)} = \infty$ follows.
\myendpf

Upon inspecting the proof of Proposition \ref{prop:Infinity-OneLaw+FirstMoment+Part1}
we see (with the help of (\ref{eq:LowerBoundFirstMomentA}))
that we have also shown the following result to be used shortly.

\begin{proposition}
{\sl
Assume $\Gamma(0) < \Gamma (1)$. With $\rho > 0$
further assume that (\ref{eq:BasicCondition+Part1}) holds.
For any $\rho$-admissible scaling
${\bf L} : \mathbb{N}_0 \rightarrow \mathbb{N}_0$, we have
\begin{eqnarray}
\lim_{n \rightarrow \infty} \bE{ I_n^{(0)}(L_n) }
=
\left \{
\begin{array}{ll}
\infty & \mbox{if $1 + \rho \ln  \Gamma(0) < 0$ } \\
         & \\
0       & \mbox{if $1 + \rho \ln  \Gamma(0) > 0$. } \\
\end{array}
\right .
\label{eq:Infinity-OneLaw+FirstMoment+Part1-ell=0}
\end{eqnarray}
}
\label{prop:Infinity-OneLaw+FirstMoment+Part1-ell=0}
\end{proposition}

The reason for this additional \lq\lq infinity-zero" law will soon become apparent as we turn next to the proof 
of Theorem \ref{thm:Zero-OneLaw+Part1}:

Let ${\bf L}: \mathbb{N}_0 \rightarrow \mathbb{N}_0$ denote a $\rho$-admissible scaling. 
Under the condition $1 + \rho \ln  \Gamma(0) > 0$,
Proposition \ref{prop:Infinity-OneLaw+FirstMoment+Part1} yields
$\lim_{n \rightarrow \infty} \bE{ I_n(L_n)} = 0$,
whence $\lim_{n \rightarrow \infty} \bP{ I_n(L_n) = 0 } = 1$ by the method of first moment,
and this establishes the one-law part of Theorem \ref{thm:Zero-OneLaw+Part1}.

In view of the second moment results of Section \ref{subsec:SecondMoments}, a straightforward 
application of the method of second moments to the count rvs (\ref{eq:Z=I}) appears problematic;
instead we focus on the related count variables
\begin{equation}
Z_n = I^{(0)}_n (L_n),
\quad n=1,2, \ldots 
\label{eq:Z=I^0}
\end{equation}

Under the condition $1 + \rho \ln  \Gamma(0) < 0$,
Proposition \ref{prop:Infinity-OneLaw+FirstMoment+Part1-ell=0} already gives the
convergence
$\lim_{n \rightarrow \infty} \bE{ I^{(0)}_n(L_n) } = \infty$.
If we were able to  establish the appropriate version of 
(\ref{eq:SecondMomentConditionAlsoZ}), namely
\begin{equation}
\limsup_{n \rightarrow \infty}   
\frac{ \bE{ \xi^{(0)}_{n,L_n}(1) \cdot  \xi^{(0)}_{n,L_n} (2) }  }{ \left ( \bE{\xi^{(0)}_{n,1}(L_n)  } \right )^2 }
\leq 1,
\label{eq:SecondMomentConditionAlsoForEll=0}
\end{equation}
we would then be in a position to conclude
$\lim_{n \rightarrow \infty}  \bP{ I^{(0)}_n(L_n) = 0 } = 0$
by the method of second moment applied to the rvs (\ref{eq:Z=I^0}).
Using the bound (\ref{eq:EasyBounds}) (with $\ell=0$) we would immediately obtain
$ \lim_{n \rightarrow \infty}  \bP{ I_n(L_n) = 0 } = 0$,
and the proof of the zero-law part of Theorem \ref{thm:Zero-OneLaw+Part1} would be completed.

To establish (\ref{eq:SecondMomentConditionAlsoForEll=0}) we proceed as follows:
Fix $n=2,3, \ldots $ and $L=1, \ldots $. Applying (\ref{eq:EvalFirstMomentWithEll})
(with $\ell=0$) gives
\[
\bE{ \xi^{(0)}_{n,L} (1) }
=  \left ( 1 -  \Gamma(0)^{L} \right )^{n-1} \cdot  \bP{ S_L(1)  = 0}
= \left ( 1 -  \Gamma(0)^{L} \right )^{n-1} \cdot \mu(0)^L .
\]
On the other hand,  specializing (\ref{eq:BoundOnExpectedCrossProduct}) to $k=\ell=0$ we obtain the bound
\[
\bE{  \xi^{(0)}_{n,L} (1) \cdot  \xi^{(0)}_{n,L} (2) } 
\leq
\bP{ S_L(1)  = 0} \bP{ S_L(2)  = 0}
= 
\mu(0)^{2L},
\]
whence
\[
\frac{ \bE{ \xi^{(0)}_{n,L}(1)  \cdot  \xi^{(0)}_{n,L} (2)  }  }{ \left ( \bE{\xi^{(0)}_{n,L} (1) } \right )^2 }
\leq
\frac{ \mu(0)^{2L} }
        { \left (  \left ( 1 -  \Gamma(0)^{L} \right )^{n-1} \cdot \mu(0)^L \right )^2 }
=
\frac{ 1 }
        { \left ( 1 -  \Gamma(0)^{L} \right )^{2(n-1)} }.
\]

As we substitute according to the $\rho$-admissible scaling ${\bf L}: \mathbb{N}_0 \rightarrow \mathbb{N}_0$ in this
last inequality we obtain
\begin{eqnarray}
\frac{ \bE{ \xi^{(0)}_{n,L_n}(1) \cdot  \xi^{(0)}_{n,L_n} (2) }  }{ \left ( \bE{\xi^{(0)}_{n,L_n}(1)  } \right )^2 }
\leq
\frac{ 1 }
        { \left ( 1 -  \Gamma(0)^{L_n} \right )^{2(n-1)} },
\quad n=2,3, \ldots
\nonumber
\end{eqnarray}
Let $n$ go infinity in this last inequality: Under the condition $1 + \rho \ln   \Gamma(0)  < 0$
we readily get (\ref{eq:SecondMomentConditionAlsoForEll=0}) as desired
since $\lim_{n \rightarrow \infty} \left ( 1 -  \Gamma(0)^{L_n} \right )^{n} = 1 $ 
by virtue of Lemma \ref{lem:TechnicalA} (with $\nu_n = 0$ for all $n=1,2, \ldots $).
\myendpf

The remainder of the paper deals with the proof of Theorem \ref{thm:Zero-OneLaw+Part2}.

\section{Auxiliary zero-infinity laws associated with Theorem \ref{thm:Zero-OneLaw+Part2}}
\label{sec:AuxiliaryProofThmPart2}

Although the arguments for proving Theorem \ref{thm:Zero-OneLaw+Part2} are similar to the ones
used in the proof of Theorem \ref{thm:Zero-OneLaw+Part1}, there are major differences in some of the technical details.
This should already be apparent from 
Proposition \ref{prop:COMPLEMENT_Infinity-OneLaw+FirstMoment+Part2} below which will act 
as the appropriate analog to Proposition \ref{prop:Infinity-OneLaw+FirstMoment+Part1-ell=0}.

Again we begin by investigating the appropriate first moment conditions (\ref{eq:FirstMomentConditionZ})
and (\ref{eq:FirstMomentConditionAlsoZ}). This is contained in the following
\lq\lq zero-infinity" law for the first moment -- Note the analogy with Theorem \ref{thm:Zero-OneLaw+Part2}.

\begin{proposition}
{\sl
Assume $\Gamma(0) < \Gamma (1)$. With $\rho > 0$
further assume that (\ref{eq:BasicCondition+Part2}) holds.
For any $\rho$-admissble scaling
${\bf L}: \mathbb{N}_0 \rightarrow \mathbb{N}_0$, we have
\begin{eqnarray}
\lim_{n \rightarrow \infty} \bE{ I_n(L_n) }
=
\left \{
\begin{array}{ll}
\infty & \mbox{if $1 + \rho \ln  \left ( \Gamma(1)^{\nu_\star(\rho) } \Gamma(0)^{1-\nu_\star(\rho)} \right ) < 0$ } \\
         & \\
0       & \mbox{if $1 + \rho \ln  \left ( \Gamma(1)^{\nu_\star(\rho)} \Gamma(0)^{1-\nu_\star(\rho)} \right ) > 0$} \\
\end{array}
\right .
\label{eq:Infinity-OneLaw+FirstMoment+Part2}
\end{eqnarray}
where $\nu_\star (\rho) $ is the unique solution in the interval $(0, \mu(1))$ to the equation
(\ref{eq:EqnForNU}).
}
\label{prop:Infinity-OneLaw+FirstMoment+Part2}
\end{proposition}

We give two proofs of Proposition \ref{prop:Infinity-OneLaw+FirstMoment+Part2}.
The first one is given in Section \ref{sec:Proof+Infinity-OneLaw+FirstMoment+Part2}
and uses Stirling's approximation to obtain the asymptotic of various quantities.
The second proof is given in Appendix
(Section \ref{app:ZeroLawPart2} and Section \ref{app:InfinityLawPart2}), 
and relies on a change of measure argument introduced in Section \ref{app:ChangeMeasure}. 
While this second  proof might be less intuitive than the one provided in this section, 
it has the advantage of giving a probabilistic interpretation to the quantity (\ref{eq:DefnG}).

As in the proof Theorem \ref{thm:Zero-OneLaw+Part1} we need to complement the \lq\lq zero-infinity" law
of Proposition \ref{prop:Infinity-OneLaw+FirstMoment+Part2}. This time, however, 
the needed result assumes a more complicated form than the one taken in
Proposition \ref{prop:Infinity-OneLaw+FirstMoment+Part1-ell=0}. 
First we need to set the stage:
Our starting point is a scaling ${\bf L}: \mathbb{N}_0 \rightarrow \mathbb{N}_0$  with the property
$\lim_{n \rightarrow \infty} L_n = \infty$, a condition
automatically satisfied by $\rho$-admissible scalings.
Pick $\nu$ in $(0,1)$, and consider any sequence
$\myvec{\ell}: \mathbb{N}_0 \rightarrow \mathbb{N}$ such that
\begin{equation}
\ell_n \leq  L_n,
\quad n=1,2, \ldots 
\label{eq:ellLessThanL}
\end{equation}
under the additional property
\begin{equation}
\lim_{n \rightarrow \infty} \frac{ \ell_n}{L_n} = \nu .
\label{eq:Convergence ToNU}
\end{equation}
We refer to any sequence $\myvec{\ell} : \mathbb{N}_0 \rightarrow \mathbb{N}$ satisfying the conditions
(\ref{eq:ellLessThanL})-(\ref{eq:Convergence ToNU})
as a sequence $\nu$-{\em associated} with the scaling ${\bf L}: \mathbb{N}_0 \rightarrow \mathbb{N}_0$.
A $\nu$-associated sequence can be easily generated through the formula
$\ell_n = \lfloor \nu L_n \rfloor$ for all $n=1,2, \ldots $.

Any $\nu$-associated sequence 
$\myvec{\ell}: \mathbb{N}_0 \rightarrow \mathbb{N}$ induces the sequence 
$\myvec{\nu}: \mathbb{N}_0 \rightarrow [0,1]$ defined by
\[
\nu_n = \frac{\ell_n}{L_n},
\quad n=1,2, \ldots 
\]
In this notation the constraints (\ref{eq:ellLessThanL})  and (\ref{eq:Convergence ToNU}) can now be expressed as 
\begin{equation}
\ell_n = \nu_n  L_n,
\quad n=1,2, \ldots 
\label{eq:ellLessThanL2}
\end{equation}
and
\begin{equation}
\lim_{n \rightarrow \infty} \nu_n = \nu .
\label{eq:Convergence ToNU2}
\end{equation}
The next result is established in Section \ref{sec:Proof+COMPLEMENT_Infinity-OneLaw+FirstMoment+Part2}.

\begin{proposition}
{\sl
Assume $\Gamma(0) < \Gamma (1)$. With $\rho > 0$, we further assume that (\ref{eq:BasicCondition+Part2}) holds.
Consider an $\rho$-admissible scaling  ${\bf L}: \mathbb{N}_0 \rightarrow \mathbb{N}_0$, 
and any $\nu$-associated sequence 
$\myvec{\ell}: \mathbb{N}_0 \rightarrow \mathbb{N}$ with $\nu$ in $(0,1)$.
Under the condition (\ref{eq:ConditionForNUstar-}), the parameter $\nu$ can be selected in the interval $(\nu_\star(\rho),\mu(1))$
so that
\begin{eqnarray}
\lim_{n \rightarrow \infty} \bE{ I^{(\ell_n)}_n(L_n) }
=
\infty .
\label{eq:COMPLEMENT_Infinity-OneLaw+FirstMoment+Part2}
\end{eqnarray}
}
\label{prop:COMPLEMENT_Infinity-OneLaw+FirstMoment+Part2}
\end{proposition}

In Section \ref{sec:Proof+Infinity-OneLaw+FirstMoment+Part2}
and Section \ref{sec:Proof+COMPLEMENT_Infinity-OneLaw+FirstMoment+Part2} 
we will have the opportunity 
to use Stirling's approximation for factorials given by
\begin{equation}
p! \sim \sqrt{2\pi p} \left (\frac{p}{e}  \right )^p
\quad (p \rightarrow \infty).
\label{eq:Stirling}
\end{equation}

\section{A proof of  Theorem \ref{thm:Zero-OneLaw+Part2} }
\label{sec:CompleteProofThmPart2}

Consider a $\rho$-admissible scaling ${\bf L}: \mathbb{N}_0 \rightarrow \mathbb{N}_0$ for some $\rho > 0$.

Under the condition 
$1 + \rho \ln  \Gamma(1)^{\nu_\star(\rho)} \Gamma(0)^{1-\nu_\star(\rho)} > 0$,
Proposition \ref{prop:Infinity-OneLaw+FirstMoment+Part2} yields
$\lim_{n \rightarrow \infty} \bE{ I_n(L_n)} = 0$,
whence $\lim_{n \rightarrow \infty} \bP{ I_n(L_n) = 0 } = 1$ 
by the method of first moments,
and this establishes the one-law part of Theorem \ref{thm:Zero-OneLaw+Part2}.
\myendpf

Assume now that 
$1 + \rho \ln  \Gamma(1)^{\nu_\star(\rho)} \Gamma(0)^{1-\nu_\star(\rho)} < 0$.
Here as well, we will not attempt to apply the method of second moment directly to the count variables (\ref{eq:Z=I}) in order
to establish the zero-law part of Theorem \ref{thm:Zero-OneLaw+Part2}. 
Under the enforced assumptions, we shall show instead 
that the parameter $\nu$ can be selected in $(\nu_\star(\rho), \mu(1)))$ 
in such a manner that the method of second moment applies 
to the count variables
\begin{equation}
Z_n = I^{(\ell_n)}_n (L_n),
\quad n=1,2, \ldots 
\label{eq:Z=I^l_Ell_n}
\end{equation}
where the sequence 
$\myvec{\ell}: \mathbb{N}_0 \rightarrow \mathbb{N}$ is $\nu$-associated 
with the scaling ${\bf L}: \mathbb{N}_0 \rightarrow \mathbb{N}_0$ for the selected value of $\nu$.

This will require showing the validity of both
\begin{equation}
\lim_{n \rightarrow \infty} 
\bE{   I^{(\ell_n)}_n (L_n) } = \infty
\label{eq:FirstMomentConditionForEll_n}
\end{equation}
and
\begin{equation}
\limsup_{n \rightarrow \infty}
\frac{ \bE{ \xi^{(\ell_n)}_{n,L_n}(1) \cdot  \xi^{(\ell_n)}_{n,L_n} (2) }  }{ \left ( \bE{\xi^{(\ell_n)}_{n,L_n}(1)  } \right )^2 }
\leq 1.
\label{eq:SecondMomentConditionAlsoForEll_n}
\end{equation}
Once this is done, it will follow from the method of second moment
applied to the rvs (\ref{eq:Z=I^l_Ell_n}) that
$\lim_{n \rightarrow \infty}  \bP{ I^{(\ell_n)}_n(L_n) = 0 } = 0$.
Using the bound (\ref{eq:EasyBounds}) (with $L=L_n$ 
and $\ell=\ell_n$ for each $n=2,3, \ldots $) we immediately obtain
$ \lim_{n \rightarrow \infty}  \bP{ I_n(L_n) = 0 } = 0$,
and the zero-law part of Theorem \ref{thm:Zero-OneLaw+Part2} will then be established.

To establish the convergence statements
(\ref{eq:FirstMomentConditionForEll_n}) and (\ref{eq:SecondMomentConditionAlsoForEll_n}), 
we proceed as follows:
By Proposition \ref{prop:COMPLEMENT_Infinity-OneLaw+FirstMoment+Part2} we already know that
there exists some $\nu$ in the interval $(\nu_\star(\rho),\mu(1))$ such that
(\ref{eq:COMPLEMENT_Infinity-OneLaw+FirstMoment+Part2}), namely (\ref{eq:FirstMomentConditionForEll_n}),
holds -- In fact the proof shows that it happens for $\nu$ in the interval $(\nu_\star(\rho), \beta_-(\rho))$.
It remains only to establish (\ref{eq:SecondMomentConditionAlsoForEll_n}) for any $\nu$ selected in 
the interval $(\nu_\star(\rho), \beta_-(\rho))$.
To that end, fix $n=2,3, \ldots $ and $L =1,2, \ldots $. Using the expression (\ref{eq:EvalFirstMomentWithEll}) we obtain
\begin{eqnarray}
\bE{ \xi^{(\ell)}_{n,L} (1) }
&=& \left ( 1 -  \Gamma(1)^{\ell} \Gamma(0)^{L-\ell} \right )^{n-1} \cdot  \bP{ S_L(1)  = \ell }
\nonumber \\
&=&
\left ( 1 -  \Gamma(1)^{\ell} \Gamma(0)^{L-\ell} \right )^{n-1}  \cdot { L \choose \ell } \mu(1)^\ell \mu(0)^{L-\ell}
\end{eqnarray}
on the range $\ell = 0,1, \ldots , L$, 
On the other hand,  specializing (\ref{eq:BoundOnExpectedCrossProduct}) to $k=\ell$ yields
\begin{eqnarray}
\bE{  \xi^{(\ell)}_{n,L} (1) \cdot  \xi^{(\ell)}_{n,L} (2) } 
\leq
\bP{ S_L(1)  = \ell } \bP{ S_L(2)  = \ell}
=
 \left ( { L \choose \ell } \mu(1)^\ell \mu(0)^{L-\ell} \right )^2
\end{eqnarray}
whence
\begin{eqnarray}
\frac{ \bE{ \xi^{(\ell)}_{n,L}(1)  \cdot  \xi^{(\ell)}_{n,L} (2)  }  }{ \left ( \bE{\xi^{(\ell)}_{n,1} (L) } \right )^2 }
&\leq&
\left ( 1 -  \Gamma(1)^{\ell} \Gamma(0)^{L-\ell} \right )^{-2(n-1)} .
\nonumber
\end{eqnarray}

Now, substitute in this last inequality according to the given $\rho$-admissible scaling ${\bf L}: \mathbb{N}_0 \rightarrow \mathbb{N}_0$
and the sequence $\myvec{\ell}: \mathbb{N}_0 \rightarrow \mathbb{N}$ $\nu$-associated with it 
where $\nu$ appearing in (\ref{eq:Convergence ToNU2}) is the one selected earlier 
in the interval $(\nu_\star(\rho), \beta_-(\rho))$.
This yields
\begin{eqnarray}
\frac{ \bE{ \xi^{(\ell_n)}_{n,L_n}(1) \cdot  \xi^{(\ell_n)}_{n,L_n} (2) }  }{ \left ( \bE{\xi^{(\ell_n)}_{n,L_n}(1)  } \right )^2 }
&\leq&
        \left ( 1 -  \Gamma(1)^{\ell_n} \Gamma(\ell_n)^{L-\ell_n} \right )^{-2(n-1)} 
\nonumber \\
&=&
        \left ( \left ( 1 -  \left ( \Gamma(1)^{\nu_n} \Gamma(0)^{(1-\nu_n)} \right )^{L_n} \right )^{(n-1)}  \right )^{-2},
\quad n=2,3, \ldots
\nonumber 
\end{eqnarray}
Letting $n$ go infinity in this last inequality we conclude 
$\lim_{n \rightarrow \infty}  \left ( 1 -  \left ( \Gamma(1)^{\nu_n} \Gamma(0)^{(1-\nu_n)} \right )^{L_n} \right )^{n-1} = 1 $
by virtue of Lemma \ref{lem:TechnicalA} 
since $1 + \rho \ln   \Gamma(1)^{\nu} \Gamma(0)^{1-\nu}  < 0$ for the value $\nu$ we selected
in the interval $(\nu_\star(\rho), \beta_-(\rho))$.
This establishes (\ref{eq:SecondMomentConditionAlsoForEll_n}) and the proof of Theorem \ref{thm:Zero-OneLaw+Part2}
is now complete.
\myendpf

\section{A proof of Proposition \ref{prop:COMPLEMENT_Infinity-OneLaw+FirstMoment+Part2}}
\label{sec:Proof+COMPLEMENT_Infinity-OneLaw+FirstMoment+Part2}

Fix $n=2,3, \ldots $ and $L=1,2, \ldots $. 
Our point of departure is the expression (\ref{eq:IwithEll}), namely
\begin{eqnarray}
\bE{ I^{(\ell)}_{n} (L) }
&=& n \left ( 1 -  \Gamma(1)^{\ell} \Gamma(0)^{L-\ell} \right )^{n-1} 
\cdot  \bP{ S_L = \ell }
\nonumber \\
&=& n \left ( 1 -  \Gamma(1)^{\ell} \Gamma(0)^{L-\ell} \right )^{n-1}  
\cdot { L \choose \ell } \mu(1)^\ell \mu(0)^{L-\ell}
\nonumber
\end{eqnarray}
on the range $\ell=0,1, \ldots , L$.

Pick $\nu$ in $(0,1)$.
Substituting $L$ and $\ell$ in this last relation according to the scaling
$\myvec{L}: \mathbb{N}_0 \rightarrow \mathbb{N}_0$ 
and any $\nu$-associated sequence 
$\myvec{\ell}: \mathbb{N}_0 \rightarrow \mathbb{N}$ 
satisfying (\ref{eq:ellLessThanL}) (or equivalently, (\ref{eq:ellLessThanL2}))
and (\ref{eq:Convergence ToNU}) for the selected $\nu$, we get
\begin{eqnarray}
\bE{ I^{(\ell_n)}_{n} (L_n) }
&=&
n  \left ( 1 -  \Gamma(1)^{\ell_n} \Gamma(0)^{L_n-\ell_n} \right )^{n-1}  
\cdot { L_n \choose \ell_n} \mu(1)^{\ell_n} \mu(0)^{L_n-\ell_n}
\nonumber \\
&=&
n { L_n \choose \nu_n L_n} \left ( \mu(1)^{\nu_n} \mu(0)^{1-\nu_n}  \right )^{L_n}
\cdot  \left ( 1 -  \left ( \Gamma(1)^{\nu_n} \Gamma(0)^{1-\nu_n} \right )^{L_n} \right )^{n-1} 
\nonumber
\end{eqnarray}
where $\nu_n L_n$ and $L_n - \nu_n L_n = (1-\nu_n)L_n$ are integers by construction.

After standard simplifications, Stirling's formula readily yields
\begin{eqnarray}
{ L_n \choose \nu_n L_n}
&\sim&
\frac{ 
\sqrt{2\pi L_n} \left ( L_n  \right )^{L_n}
}
{
\sqrt{2\pi \nu_n L_n} \left ( \nu_n L_n  \right )^{\nu_n L_n} 
\cdot 
\sqrt{2\pi (1-\nu_n) L_n} \left ( (1-\nu_n)L_n  \right )^{(1-\nu_n)L_n}
}
\nonumber \\
&=& \frac{1}
                 { \sqrt{ 2 \pi \nu_n (1-\nu_n) L_n} }
        \cdot
        \frac{1}
                { \left ( \nu_n^{\nu_n} (1-\nu_n)^{1-\nu_n} \right )^{L_n}}
\nonumber
\end{eqnarray}
so that
\begin{eqnarray}
n { L_n \choose \nu_n L_n} \left ( \mu(1)^{\nu_n} \mu(0)^{1-\nu_n}  \right )^{L_n}
&\sim&
\frac{n}
                 { \sqrt{ 2 \pi \nu_n (1-\nu_n) L_n} }
        \cdot
        \left ( \frac{ \mu(1)^{\nu_n} \mu(0)^{1-\nu_n} }
                          { \nu_n^{\nu_n} (1-\nu_n)^{1-\nu_n} }
        \right )^{L_n}
\nonumber \\
&=& \frac{n}
                 { \sqrt{ 2 \pi \nu_n (1-\nu_n) L_n} }
        \cdot G( \nu_n , \mu(1) )^{L_n}.
\end{eqnarray}
Collecting we obtain
\begin{eqnarray}
\bE{ I^{(\ell_n)}_{n} (L_n) }
&\sim&
 \frac{n}
                 { \sqrt{ 2 \pi \nu_n (1-\nu_n) L_n} }
        \cdot G( \nu_n , \mu(1) )^{L_n}
\cdot
 \left ( 1 -  \left ( \Gamma(1)^{\nu_n} \Gamma(0)^{1-\nu_n} \right )^{L_n} \right )^{n-1} 
 \nonumber \\
&\sim&
\frac{1}
                 { \sqrt{ 2 \pi \nu (1-\nu)} }
        \cdot \frac{n \cdot G( \nu_n , \mu(1) )^{L_n} }{\sqrt{L_n}}
\cdot
 \left ( 1 -  \left ( \Gamma(1)^{\nu_n} \Gamma(0)^{1-\nu_n} \right )^{L_n} \right )^{n-1} 
 \nonumber
\end{eqnarray}
as we make use of (\ref{eq:Convergence ToNU2}) in the last step.

Recall now that both conditions (\ref{eq:BasicCondition+Part2})
and (\ref{eq:ConditionForNUstar-}) are enforced.
Therefore, as discussed at the end of Section \ref{sec:Results},
condition (\ref{eq:ConditionForNU-}) holds on some interval $I_-(\rho) = ( \alpha_-(\rho) , \beta_-(\rho) ) \subseteq (0, \mu(1))$,
said interval containing $\nu_\star(\rho)$.
As we {\em restrict} $\nu$ to be an element of $(\nu_\star(\rho), \beta_-(\rho) )$, we conclude
by Lemma \ref{lem:TechnicalA} that
\begin{equation}
\lim_{n \rightarrow \infty}  \left ( 1 -  \left ( \Gamma(1)^{\nu_n} \Gamma(0)^{1-\nu_n} \right )^{L_n} \right )^{n-1}  = 1,
\end{equation}
and the desired conclusion 
$\lim_{n \rightarrow \infty}  \bE{ I^{(\ell_n)}_{n} (L_n) } = \infty$ follows provided we can show that
\begin{equation}
\liminf_{n \rightarrow \infty}  \frac{n \cdot G( \nu_n , \mu(1) )^{L_n} }{\sqrt{L_n}} > 0.
\label{eq:LimInfCondition}
\end{equation}

It is always possible to find $\varepsilon > 0$ so that the interval 
$ (\nu - \varepsilon ,  \nu + \varepsilon )$ is contained in the interval $(\nu_\star(\rho), \beta_-(\rho) )$.
By virtue of (\ref{eq:Convergence ToNU}) there exists a finite integer $n(\varepsilon)$ such that
$\nu - \varepsilon < \nu_n <  \nu + \varepsilon$ whenever $n \geq n(\varepsilon)$,
and on that range, the monotonicity of the mapping $\nu^\prime \rightarrow 1 + \rho \ln  G( \nu^\prime , \mu(1) ) $ on $(0, \mu(1))$
yields
\[
0 
<
1+ \rho \ln  G( \nu - \varepsilon  , \mu(1) )
\leq
1+ \rho \ln  G( \nu_n , \mu(1) )
\]
because $1+ \rho \ln  G( \nu^\prime  , \mu(1) ) > 0$ on the interval $(\nu_\star(\rho), \beta_-(\rho) )$.
Returning to the proof of Lemma \ref{lem:TechnicalB} (with $C_n = G( \nu_n , \mu(1) )$ for all $n=1,2, \ldots$),
we see that (\ref{eq:TechnicalB_bound}) yields the bounds
\begin{eqnarray}
n \cdot G( \nu_n , \mu(1) )^{L_n} 
= n^{1 + \rho_n \ln  G( \nu_n , \mu(1) ) }
\geq
n^{1 + \rho_n \ln  G( \nu - \varepsilon , \mu(1) ) },
\quad n \geq n(\varepsilon) 
\nonumber
\end{eqnarray}
where the sequence $\myvec{\rho}: \mathbb{N}_0 \rightarrow \mathbb{R}_+$ is
the unique sequence associated with the $\rho$-admissible scaling $\myvec{L}: \mathbb{N}_0 \rightarrow \mathbb{N}_0$.
It is then plain that
\begin{eqnarray}
\liminf_{n \rightarrow \infty}  \frac{n \cdot G( \nu_n , \mu(1) )^{L_n} }{\sqrt{L_n}} 
&\geq&
\liminf_{n \rightarrow \infty}  \frac{ n^{1 + \rho_n \ln  G( \nu - \varepsilon , \mu(1) ) } }
                                                          {\sqrt{ \rho_n \ln  n } }
                                                          = \infty 
\end{eqnarray}
since $1+ \rho \ln  G( \nu - \varepsilon  , \mu(1) ) > 0$.
This establishes (\ref{eq:LimInfCondition}), 
and the proof of  Proposition \ref{prop:COMPLEMENT_Infinity-OneLaw+FirstMoment+Part2}
is now complete.
\myendpf

\section{A proof of Proposition \ref{prop:Infinity-OneLaw+FirstMoment+Part2}}
\label{sec:Proof+Infinity-OneLaw+FirstMoment+Part2}

Assume $\Gamma(0) < \Gamma (1)$, and 
consider a $\rho$-admissible scaling ${\bf L}: \mathbb{N}_0 \rightarrow \mathbb{N}_0$  for some $\rho > 0$. 

Under the condition  $1 + \rho \ln  \Gamma(1)^{\nu_\star(\rho)} \Gamma(0)^{1-\nu_\star(\rho)} <  0$,
Proposition \ref{prop:COMPLEMENT_Infinity-OneLaw+FirstMoment+Part2} 
asserts the existence of $\nu$ in $(0,1)$ such that $\lim_{n \rightarrow \infty} \bE{ I^{(\ell_n)} _n(L_n)} = \infty$
for any $\nu$-associated sequence $\myvec{\ell}: \mathbb{N}_0 \rightarrow \mathbb{N}$.
It now follows that $\lim_{n \rightarrow \infty} \bE{ I_n(L_n)} = \infty$, and the infinity part of 
Proposition \ref{prop:Infinity-OneLaw+FirstMoment+Part2} holds -- This is an immediate consequence 
of the bound (\ref{eq:EasyBounds}) (with $L=L_n$  and $\ell=\ell_n$ for each $n=2,3, \ldots $).
\myendpf

As we now turn to establishing the zero-law in (\ref{eq:Infinity-OneLaw+FirstMoment+Part2}), assume that
the condition  $1 + \rho \ln  \Gamma(1)^{\nu_\star(\rho)} \Gamma(0)^{1-\nu_\star(\rho)} > 0$ holds:
As discussed at the end of Section \ref{sec:Results},  under this condition  
there exists $\varepsilon $ sufficiently small in $(0,\nu_\star(\rho))$ so that $\alpha_+(\rho) < \nu_\star (\rho) - \varepsilon$,
hence $1 + \rho \ln \left ( \Gamma(1)^{\nu_\star(\rho)-\varepsilon} \Gamma(0)^{1-\nu_\star(\rho)+\varepsilon} \right ) > 0$.
Select such a value of $\varepsilon$ and keep it fixed throughout the proof.

Fix $n=2,3,\dots$. It follows from (\ref{eq:I}) that  
\begin{eqnarray}
\bE{ I_{n} (L_n) }
&=& n\sum_{\ell=0}^{L_n}\binom{L_n}{\ell}\mu(1)^{\ell}\mu(0)^{L_n-\ell}\left ( 1 - \Gamma(1)^{\ell} \Gamma(0)^{L_n-\ell} \right )^{n-1}
\nonumber\\
&=& \sum_{\ell=0}^{\lfloor (\nu_\star(\rho)-\varepsilon) L_n \rfloor}
n\binom{L_n}{\ell}\mu(1)^{\ell}\mu(0)^{L_n-\ell}\left ( 1 - \Gamma(1)^{\ell} \Gamma(0)^{L_n-\ell} \right )^{n-1} 
\nonumber \\
& & ~+  \sum_{\ell=\lfloor (\nu_\star(\rho)-\varepsilon) L_n \rfloor +1}^{L_n}
n\binom{L_n}{\ell}\mu(1)^{\ell}\mu(0)^{L_n-\ell}\left ( 1 - \Gamma(1)^{\ell} \Gamma(0)^{L_n-\ell} \right )^{n-1}.
\nonumber
\end{eqnarray}
We will obtain the desired conclusion $\lim_{n \rightarrow \infty} \bE{ I_n(L_n)} = 0$ by showing that
\begin{equation}
\lim_{n \rightarrow \infty} 
\sum_{\ell=0}^{\lfloor (\nu_\star(\rho)-\varepsilon) L_n \rfloor}
n\binom{L_n}{\ell}\mu(1)^{\ell}\mu(0)^{L_n-\ell}\left ( 1 - \Gamma(1)^{\ell} \Gamma(0)^{L_n-\ell} \right )^{n-1} 
= 0
\label{eq:Limit+1} 
\end{equation}
and
\begin{equation}
\lim_{n \rightarrow \infty}  
\sum_{\ell=\lfloor (\nu_\star(\rho)-\varepsilon) L_n \rfloor +1}^{L_n}
n\binom{L_n}{\ell}\mu(1)^{\ell}\mu(0)^{L_n-\ell}\left ( 1 - \Gamma(1)^{\ell} \Gamma(0)^{L_n-\ell} \right )^{n-1} 
= 0.
 \label{eq:Limit+2}
\end{equation}

To establish (\ref{eq:Limit+1}) we proceed as follows:
First, for $\ell =0,1,\dots, L_n$, note the crude bounds
\begin{eqnarray*}
n\binom{L_n}{\ell}\mu(1)^{\ell}\mu(0)^{L_n-\ell}\left ( 1 - \Gamma(1)^{\ell} \Gamma(0)^{L_n-\ell} \right )^{n-1} 
\leq  n\binom{L_n}{\ell}\mu(1)^{\ell}\mu(0)^{L_n-\ell}.
\end{eqnarray*}
Since $\nu_\star(\rho)-\varepsilon$ lies in $(0,\mu(1))$, the quantity $\binom{L_n}{\ell}\mu(1)^{\ell}\mu(0)^{L_n-\ell}$ 
increases with $\ell$ on the range $\ell = 0, 1, \ldots , \lfloor (\nu_\star(\rho)-\varepsilon)L_n \rfloor$, 
and we obtain the bound
\begin{eqnarray}
\lefteqn{\sum_{\ell=0}^{\lfloor (\nu_\star(\rho)-\varepsilon) L_n \rfloor}
n\binom{L_n}{\ell}\mu(1)^{\ell}\mu(0)^{L_n-\ell}\left ( 1 - \Gamma(1)^{\ell} \Gamma(0)^{L_n-\ell} \right )^{n-1}}\nonumber\\
&\leq&  
L_n \cdot n  
\binom{L_n}{\lfloor (\nu_\star(\rho)-\varepsilon) L_n \rfloor}\mu(1)^{\lfloor (\nu_\star(\rho)-\varepsilon) L_n \rfloor}\mu(0)^{L_n-\lfloor (\nu_\star(\rho)-\varepsilon) L_n \rfloor}.
\label{eq:Part2ZeroInftyLower}
\end{eqnarray}

Using Stirling's formula, we get the asymptotic equivalence
\begin{eqnarray}
\binom{L_n}{\lfloor (\nu_\star(\rho)-\varepsilon) L_n \rfloor}
\sim
\frac{ \sqrt{L_n} }
        { \sqrt{ 2 \pi \lfloor (\nu_\star(\rho)-\varepsilon) L_n \rfloor \cdot ( L_n - \lfloor (\nu_\star(\rho)-\varepsilon) L_n \rfloor ) }
        } 
        \cdot A_n
\end{eqnarray}
where for each $n=1,2, \ldots $, the factor $A_n$ is given by
\begin{eqnarray}
A_n
&\equiv&
\left ( \frac{ L_n}{ \lfloor (\nu_\star(\rho)-\varepsilon) L_n \rfloor } \right )^{ \lfloor (\nu_\star(\rho)-\varepsilon) L_n \rfloor }
\cdot
\left ( \frac{ L_n}{ L_n - \lfloor (\nu_\star(\rho)-\varepsilon) L_n \rfloor } \right )^{ L_n - \lfloor (\nu_\star(\rho)-\varepsilon) L_n \rfloor }.
\nonumber
\end{eqnarray}
After simplifications and rearrangements it follows that
\begin{eqnarray}
\lefteqn{
\binom{L_n}{\lfloor (\nu_\star(\rho)-\varepsilon) L_n \rfloor}
\cdot \mu(1)^{\lfloor (\nu_\star(\rho)-\varepsilon) L_n \rfloor}\mu(0)^{L_n-\lfloor (\nu_\star(\rho)-\varepsilon) L_n \rfloor}
} & &
\nonumber \\
&\sim&
\frac{ \sqrt{L_n} }
        { \sqrt{ 2 \pi \lfloor (\nu_\star(\rho)-\varepsilon) L_n \rfloor \cdot ( L_n - \lfloor (\nu_\star(\rho)-\varepsilon) L_n \rfloor ) }
        } 
        \cdot A^\star_n
\end{eqnarray}
where for each $n=1,2, \ldots $ we have
\begin{eqnarray}
A^\star_n
&\equiv&
\left ( \frac{ \mu(1) L_n}{ \lfloor (\nu_\star(\rho)-\varepsilon) L_n \rfloor } \right )^{ \lfloor (\nu_\star(\rho)-\varepsilon) L_n \rfloor }
\cdot
\left ( \frac{ \mu(0) L_n}{ L_n - \lfloor (\nu_\star(\rho)-\varepsilon) L_n \rfloor } \right )^{ L_n - \lfloor (\nu_\star(\rho)-\varepsilon) L_n \rfloor }
\nonumber \\
&=&
G\left(\frac{\lfloor (\nu_\star(\rho)-\varepsilon) L_n \rfloor}{L_n},\mu(1)\right)^{L_n}
\end{eqnarray}
as we recall the definition (\ref{eq:DefnG}) of of $G(\cdot,\cdot)$.

Noting that
\[
\lim_{n \rightarrow \infty}
\frac{ \sqrt{L_n} }
        { \sqrt{ 2 \pi \lfloor (\nu_\star(\rho)-\varepsilon) L_n \rfloor \cdot ( L_n - \lfloor (\nu_\star(\rho)-\varepsilon) L_n \rfloor ) }
        } 
        = 0,
\]
we conclude that
\[
\binom{L_n}{\lfloor (\nu_\star(\rho)-\varepsilon) L_n \rfloor}
\cdot \mu(1)^{\lfloor (\nu_\star(\rho)-\varepsilon) L_n \rfloor}\mu(0)^{L_n-\lfloor (\nu_\star(\rho)-\varepsilon) L_n \rfloor}
< 
G\left(\frac{\lfloor (\nu_\star(\rho)-\varepsilon) L_n \rfloor}{L_n},\mu(1)\right)^{L_n}
\]
for $n$ sufficiently large, and the upper bound
\begin{eqnarray}
\lefteqn{
\sum_{\ell=0}^{\lfloor (\nu_\star(\rho)-\varepsilon) L_n \rfloor}n\binom{L_n}{\ell}\mu(1)^{\ell}\mu(0)^{L_n-\ell}\left ( 1 - \Gamma(1)^{\ell} \Gamma(0)^{L_n-\ell} \right )^{n-1}
} & &
\nonumber\\
&\leq&  L_n \cdot n  G\left(\frac{\lfloor (\nu_\star(\rho)-\varepsilon) L_n \rfloor}{L_n},\mu(1)\right)^{L_n} 
\label{eq:lowerSumBound}
\end{eqnarray}
then follows for sufficiently large $n$.

Next, the sequence $\myvec{\rho}: \mathbb{N}_0 \rightarrow \mathbb{R}_+$ being
the unique sequence associated with the $\rho$-admissible scaling $\myvec{L}: \mathbb{N}_0 \rightarrow \mathbb{N}_0$, we write
\begin{eqnarray}
L_n \cdot n  G\left(\frac{\lfloor (\nu_\star(\rho)-\varepsilon) L_n \rfloor}{L_n},\mu(1)\right)^{L_n}&=& e^{\ln (\rho_n\ln n)+  (1+\rho_n \ln C_n)\ln n} \nonumber 
\end{eqnarray}
for each $n=1,2 \ldots $ where we have set 
\[
C_n=G\left(\frac{\lfloor (\nu_\star(\rho)-\varepsilon) L_n \rfloor}{L_n},\mu(1)\right).
\]
Obviously we have $\lim\limits_{n\rightarrow \infty}\frac{\lfloor (\nu_\star(\rho)-\varepsilon) L_n \rfloor}{L_n} = \nu_\star(\rho)-\varepsilon$, 
while the definition of $\nu_\star(\rho)$ implies $1+\rho\ln G\left(\nu_\star(\rho)-\varepsilon,\mu(1)\right)<0$. Thus, 
letting $n$ go to infinity in (\ref{eq:lowerSumBound}) yields 
\begin{eqnarray*}
\lim_{n\rightarrow \infty}L_n \cdot n  G\left(\frac{\lfloor (\nu_\star(\rho)-\varepsilon) L_n \rfloor}{L_n},\mu(1)\right)^{L_n} =0
\end{eqnarray*}
and (\ref{eq:Limit+1}) holds

As we turn to showing (\ref{eq:Limit+2}) we note the successive bounds
\begin{eqnarray*}
\lefteqn{
\sum_{\ell=\lfloor (\nu_\star(\rho)-\varepsilon) L_n \rfloor+1}^{L_n} n\binom{L_n}{\ell}\mu(1)^{\ell}\mu(0)^{L_n-\ell}\left ( 1 - \Gamma(1)^{\ell} \Gamma(0)^{L_n-\ell} \right )^{n-1}
} & &
\nonumber\\ 
&\leq & 
n\left( 1- \Gamma(1)^{\lceil (\nu_\star(\rho)-\varepsilon) L_n \rceil}\Gamma(0)^{L_n-\lceil (\nu_\star(\rho)-\varepsilon) L_n \rceil}\right)^{n-1}
\nonumber\\
&\leq & 
n\left( 1- \Gamma(1)^{ (\nu_\star(\rho)-\varepsilon) L_n }\Gamma(0)^{L_n- (\nu_\star(\rho)-\varepsilon) L_n }\right)^{n-1},
\quad n=1,2, \ldots
\label{eq:Part2ZeroInftyUpper}
\end{eqnarray*}
Indeed, the quantity $\left ( 1 - \Gamma(1)^{\ell} \Gamma(0)^{L_n-\ell} \right )^{n-1}$ 
is monotonically decreasing in $\ell$ under the assumption $\Gamma(1)>\Gamma(0)$, 
and a straightforward probabilistic interpretation yields
\[
\sum_{\ell=\lfloor (\nu_\star(\rho)-\varepsilon) L_n \rfloor +1}^{L_n}\binom{L_n}{\ell}\mu(1)^{\ell}\mu(0)^{L_n-\ell}\leq 1.
\]
The condition $1 + \rho \ln \left ( \Gamma(1)^{\nu_\star(\rho)-\varepsilon} \Gamma(0)^{1-\nu_\star(\rho)+\varepsilon} \right ) > 0$ implies
\[
\lim_{n\rightarrow \infty} n\left( 1- \Gamma(1)^{ (\nu_\star(\rho)-\varepsilon) L_n }\Gamma(0)^{L_n- (\nu_\star(\rho)-\varepsilon) L_n }\right)^{n-1} = 0
\]
by the remark following the proof of Lemma \ref{lem:TechnicalA}, and
the convergence (\ref{eq:Limit+2}) holds.
This completes the proof of Proposition \ref{prop:Infinity-OneLaw+FirstMoment+Part2}
\myendpf

\section*{Acknowledgment}
This work was supported by NSF Grant CCF-1217997.
The paper was completed during the academic year 2014-2015  while A.M. Makowski 
was a Visiting Professor with the Department of Statistics of the Hebrew University of Jerusalem 
with the support of a fellowship from the Lady Davis Trust.

\bibliographystyle{IEEE}

\section{Appendix: A proof of Lemma \ref{lem:EvalSecondMoment}}
\label{app:ProofLemmaEvalSecondMoment}

The arguments are very similar to the ones given in the proof of Lemma \ref{lem:EvalFirstMoment}.
Pick positive $n=2,3, \ldots $ and $L=1,2, \ldots $, and consider distinct nodes $u,v=1, \ldots , n$.
For $k, \ell  =0,1, \ldots , L$, not necessarily distinct, we start from the relation (\ref{eq:ExpectedCrossProduct}).
Note that the product $\xi_{n,L}(u) \xi_{n,L}(v)$ can be expressed as
\begin{eqnarray}
\xi_{n,L}(u) \xi_{n,L}(v)
&=&
\prod_{w=1, \ w \neq u}^n \left ( 1 - \chi_L(u,w) \right )
\cdot
\prod_{w=1, \ w \neq v}^n \left ( 1 - \chi_L(v,w) \right )
\nonumber \\
&=&
\left ( 1 - \chi_L(u,v)\right )
\cdot
\prod_{w=1, \ w \neq u,v}^n
\left ( 1 - \chi_L(u,w) \right )  \left ( 1 - \chi_L(v,w) \right )
\nonumber
\end{eqnarray}
with factors represented as
\[
1 - \chi_L(u,v) = \1{ U(u,v) > Q_L ( \myvec{A}_L(u), \myvec{A}_L(v) ) }
\]
and
\begin{eqnarray}
&  &
\prod_{w=1, \ w \neq u,v}^n
\left ( 1 - \chi_L(u,w) \right )  \left ( 1 - \chi_L(v,w) \right )
\nonumber \\
&=&
\prod_{w=1, \ w \neq u,v}^n
\1{ U(u,w) > Q_L ( \myvec{A}_L(u), \myvec{A}_L(w) ) }  
\cdot
\1{ U(v,w) > Q_L ( \myvec{A}_L(v), \myvec{A}_L(w) ) } .
\nonumber
\end{eqnarray}
Under the enforced independence assumptions, it is now straightforward to conclude that
\begin{eqnarray}
& &
\bE{  \xi_{n,L}(u)  \xi_{n,L}(v) \Bigl |  \myvec{A}_L (1), \ldots ,  \myvec{A}_L (n) }
\nonumber \\
&=&
\left ( 1 - Q_L ( \myvec{A}_L (u), \myvec{A}_L (v) ) \right )
\cdot 
\prod_{w=1, \ w \neq u,v}^n
\left ( 1 - Q_L ( \myvec{A}_L (u), \myvec{A}_L (w) ) \right ) \left ( 1 - Q_L ( \myvec{A}_L (v), \myvec{A}_L (w) ) \right ).
\nonumber 
\end{eqnarray}
The smoothing property of conditional expectations is again invoked, this time to obtain
\begin{eqnarray}
& &
\bE{  \xi_{n,L}(u)  \xi_{n,L}(v) \Bigl |  \myvec{A}_L (u),  \myvec{A}_L (v) }
\nonumber \\
&=&
\bE{ \bE{  \xi_{n,L}(u)  \xi_{n,L}(v) \Bigl |  \myvec{A}_L (1), \ldots ,  \myvec{A}_L (n) } \Biggl |  \myvec{A}_L (u),  \myvec{A}_L (v) }
\nonumber \\
&=&
\left ( 1 - Q_L ( \myvec{A}_L (u), \myvec{A}_L (v) ) \right )
\cdot 
\bE{ \prod_{w=1, \ w \neq u,v}^n \ldots \Biggl |  \myvec{A}_L (u),  \myvec{A}_L (v) }
\label{eq:SecondMomentIntermediary}
\end{eqnarray}
where
\begin{eqnarray}
& &
\bE{  \prod_{w=1, \ w \neq u,v}^n \ldots \Bigl |  \myvec{A}_L (u),  \myvec{A}_L (v) } 
\nonumber \\
&=&
\bE{  \prod_{w=1, \ w \neq u,v}^n \left ( 1 - Q_L ( \myvec{A}_L (u), \myvec{A}_L (w) ) \right ) \left ( 1 - Q_L ( \myvec{A}_L (v), \myvec{A}_L (w) ) \right )
\Biggl |  \myvec{A}_L (u),  \myvec{A}_L (v) } 
\nonumber \\
&=&
\bE{  \prod_{w=1, \ w \neq u,v}^n \left ( 1 - Q_L ( \myvec{a}_L, \myvec{A}_L (w) ) \right ) \left ( 1 - Q_L ( \myvec{b} , \myvec{A}_L (w) ) \right )
} _{ \myvec{a}_L  =  \myvec{A}_L (u) ,  \myvec{b}_L =\myvec{A}_L (v)  }
\nonumber \\
&=&
\left ( 
\prod_{w=1, \ w \neq u,v}^n 
\bE{  \left ( 1 - Q_L ( \myvec{a}_L, \myvec{A}_L (w) ) \right ) \left ( 1 - Q_L ( \myvec{b}_L, \myvec{A}_L (w) ) \right )
}
\right ) _{ \myvec{a}_L  =  \myvec{A}_L (u) ,  \myvec{b}_L =\myvec{A}_L (v)  }
\nonumber \\
&=&
\left (
\bE{  \left ( 1 - Q_L ( \myvec{a}_L, \myvec{A}_L  ) \right ) \left ( 1 - Q_L ( \myvec{b}_L, \myvec{A}_L ) \right )
} _{ \myvec{a}_L  =  \myvec{A}_L (u) ,  \myvec{b}_L =\myvec{A}_L (v)  }
\right )^{n-2}
\end{eqnarray}
under the enforced i.i.d. assumptions on the rvs $\myvec{A}_L (1), \ldots ,  \myvec{A}_L (n) $.
In the notation introduced earlier at (\ref{eq:Q_StarL}) and (\ref{eq:Q^StarStar}) we can write
\begin{eqnarray}
\lefteqn{
\bE{
\left ( 1 - Q_L ( \myvec{a}_L , \myvec{A}_L ) \right ) \left ( 1 - Q_L ( \myvec{b}_L , \myvec{A}_L  ) \right )
}
} & &
\nonumber \\
&=&
1 -  Q^{\star}_L( \myvec{a}_L ) - Q^{\star}_L( \myvec{b}_L)  +  Q^{\star\star}_L( \myvec{a}_L, \myvec{b}_L) ,
\quad \myvec{a}_L , \myvec{b}_L  \in \{0,1\}^L.
\end{eqnarray}
This allows us to conclude that
\begin{eqnarray}
& &
\bE{  \prod_{w=1, \ w \neq u,v}^n \left ( 1 - Q_L ( \myvec{A}_L (u), \myvec{A}_L (w) ) \right ) \left ( 1 - Q_L ( \myvec{A}_L (v), \myvec{A}_L (w) ) \right )
\Biggl |  \myvec{A}_L (u),  \myvec{A}_L (v) } 
\nonumber \\
&=&
\left (
1 -  Q^{\star}_L( \myvec{A}_L(u) ) - Q^{\star}_L( \myvec{A}_L(v))  +  Q^{\star\star}_L( \myvec{A}_L(u), \myvec{A}_L (v)) 
\right )^{n-2},
\end{eqnarray}
and substituting into (\ref{eq:SecondMomentIntermediary}) we obtain 
the desired conclusion (\ref{eq:EvalSecondMomentConditional}).
\myendpf

\section{Appendix: A change of measure}
\label{app:ChangeMeasure}

As stated earlier, all rvs are defined on the measurable space $( \Omega, {\cal F} )$ and their statistics
computed under the given probability measure $\mathbb{P}$ as stipulated by Assumptions (i)-(iii).
To proceed we will find it convenient to embed $\mathbb{P}$ into 
a collection of probability measures $\{ \mathbb{P}_\nu, \ \nu \in (0,1) \}$ defined on the
$\sigma$-field ${\cal F}$ with the following properties:
For each $\nu$ in $(0,1)$, under the probability measure $\mathbb{P}_\nu$, Assumptions (i) and (ii) remain unchanged but
Assumption (iii) is replaced by the following assumption:
\begin{enumerate}
\item[(iii-$\nu$)] The rvs
$\left \{ A, A_\ell, A_\ell (u), \ \ell =1,2, \ldots ; \ u=1,2, \ldots \right \}$
form a collection of {\em i.i.d.} $\{0,1\}$-valued rvs
with pmf $\myvec{\nu} = (\nu,1-\nu)$ where
\[
\mathbb{P}_\nu [ A = 0 ] = 1-\nu \quad \mbox{and} \quad \mathbb{P}_\nu [  A = 1 ] = \nu .
\]
\end{enumerate}
Let $\mathbb{E}_\nu$ denote the expectation operator associated with $\mathbb{P}_\nu$.

Obviously, we have $\mathbb{P} \equiv  \mathbb{P}_\nu$ when selecting $\nu = \mu(1)$.
It is always possible to construct a measurable space $( \Omega, {\cal F} )$, the appropriate collections of rvs on it
and a collection $\{ \mathbb{P}_\nu, \ \nu \in (0,1)\}$ of probability measures 
defined on the $\sigma$-field ${\cal F}$ with the requisite properties; details
are well known and omitted here for the sake of brevity.

In fact, given $\nu$ in $(0,1)$, for each $L=1, \ldots$,
the probability measures $\mathbb{P}$ and $\mathbb{P}_\nu$ are mutually
absolutely continuous  when restricted to the $\sigma$-field $\sigma \{ A_1, \ldots , A_L\}$
with Radon-Nikodym derivative given by
\[
\left ( \frac{ d \mathbb{P} }{ d \mathbb{P}_\nu } \right )_{L}
= \prod_{\ell=1}^L 
\left ( \frac{\mu(1)}{\nu} \right )^{A_\ell} \left ( \frac{1-\mu(1)}{1-\nu} \right )^{1-A_\ell}
= 
\left ( \frac{\mu(1)}{\nu} \right )^{S_L} \left ( \frac{1-\mu(1)}{1-\nu} \right )^{L-S_L}.
\]
However,  the probability measures $\mathbb{P}$ and $\mathbb{P}_\nu$ 
are not mutually absolutely continuous on the entire $\sigma$-field ${\cal F}$.

To take advantage of this change of measure we proceed as follows:
Fix $\nu$ in $(0,1)$, $n=2,3, \ldots $ and $L=1,2, \ldots$. 
The expression (\ref{eq:I}) can be written
\begin{eqnarray}
\bE{I_n(L)}
&=&
n \bE{ 
\left ( 1 - \Gamma(1)^{S_L} \Gamma(0)^{L-S_L} \right )^{n-1} 
}
\nonumber \\
&=&
n \cdot
\mathbb{E}_{\nu} 
\left [
\left ( 1 - \Gamma(1)^{S_L} \Gamma(0)^{L-S_L} \right )^{n-1} 
\cdot  \left ( \frac{\mu(1)}{\nu} \right )^{S_L}  \left ( \frac{1-\mu(1)}{1-\nu} \right )^{L- S_L}
\right ]
\nonumber \\
&=&
n  \left ( \left ( \frac{\mu(1)}{\nu} \right )^{\nu}  \left ( \frac{1-\mu(1)}{1-\nu} \right )^{1-\nu} \right )^L
\cdot
E_n ( \nu ,L)
\nonumber \\
\nonumber \\
&=&
n G( \nu, \mu(1))^L  \cdot E_n ( \nu ,L)
\label{eq:ChangeMeasure1}
\end{eqnarray}
with the definition (\ref{eq:DefnG}) used in the last step and where we have set
\begin{equation}
E_n ( \nu ,L)
=
\mathbb{E}_{\nu} 
\left [
\left ( 1 - \Gamma(1)^{S_L} \Gamma(0)^{L-S_L} \right )^{n-1} 
\cdot
\left ( \frac{\mu(1)}{\nu} \cdot \frac{1-\nu}{1-\mu(1)} \right )^{ S_L - L \nu } 
\right ].
\label{eq:ChangeMeasure2}
\end{equation}

For future reference we note the decomposition
\begin{equation}
E_n ( \nu ,L) = E^+_n ( \nu ,L)  + E^-_n ( \nu ,L) 
\label{eq:Decomposition}
\end{equation}
with $E^+_n ( \nu ,L)$ and $E^-_n ( \nu ,L)$ given by
\[
E^+_n ( \nu ,L) 
= 
\mathbb{E}_{\nu} 
\left [
\left ( 1 - \Gamma(1)^{S_L} \Gamma(0)^{L-S_L} \right )^{n-1} 
\cdot
\left ( \frac{\mu(1)}{\nu} \cdot \frac{1-\nu}{1-\mu(1)} \right )^{ S_L - L \nu } 
\1{ S_L - \nu L >  0 }
\right ]
\]
and
\[
E^-_n ( \nu ,L) 
= 
\mathbb{E}_{\nu} 
\left [
\left ( 1 - \Gamma(1)^{S_L} \Gamma(0)^{L-S_L} \right )^{n-1} 
\cdot
\left ( \frac{\mu(1)}{\nu} \cdot \frac{1-\nu}{1-\mu(1)} \right )^{ S_L - L \nu } 
\1{ S_L - \nu L \leq  0 }
\right ].
\]

It is plain that
\begin{equation}
\frac{\mu(1)}{\nu} \cdot \frac{1 - \nu }{1-\mu(1)} > 1
\mbox{~if and only if~} \nu < \mu(1) .
\label{eq:BoundOnRatio}
\end{equation}
We shall also use the simple fact that
\begin{equation}
\Gamma(1)^{S_L} \Gamma(0)^{L-S_L}
= \left ( \Gamma(1)^\nu \Gamma(0)^{1-\nu} \right )^L
\cdot
\left ( \frac{\Gamma(1)}{\Gamma(0)}
\right )^{ S_L - L \nu }.
\label{eq:Ratio}
\end{equation}
These observations form the basis for the arguments given next.

\section{Appendix: A proof of Proposition \ref{prop:Infinity-OneLaw+FirstMoment+Part2} -- The zero-law}
\label{app:ZeroLawPart2}

Consider a $\rho$-admissible scaling
${\bf L}: \mathbb{N}_0 \rightarrow \mathbb{N}_0$ such that (\ref{eq:BasicCondition+Part2}) holds, or equivalently,
\begin{equation}
1 + \rho \ln  (1-\mu (1)) < 0.
\label{eq:BasicCondition+Part2Bis}
\end{equation}
By the discussion preceding the statement of Theorem \ref{thm:Zero-OneLaw+Part2} the non-linear equation
(\ref{eq:EqnForNU}) admits a single solution $\nu_\star (\rho)$ in the interval $(0, \mu(1))$  and
\[
1 + \rho \ln  G(\nu, \mu (1)) < 0 ,
\quad \nu \in (0, \nu_\star (\rho)).
\]
It follows from Lemma \ref{lem:TechnicalB} (with $C_n = G( \nu, \mu(1))$ for all $n=1,2, \ldots$) that
\[
\lim_{n \rightarrow \infty} n G( \nu, \mu(1))^{L_n}  = 0,
\quad \nu \in (0, \nu_\star (\rho)).
\quad 
\]

Therefore, by virtue of (\ref{eq:ChangeMeasure1}) the desired result 
$\lim_{n \rightarrow \infty} \bE{ I_n (L_n) } = 0$
will be established if we show that
\begin{equation}
\limsup_{n \rightarrow \infty} E_n ( \nu ,L_n ) < \infty
\label{eq:BoundedA}
\end{equation}
for {\em some} $\nu$ in $(0, \nu_\star (\rho))$.

This issue is explored with the help of the decomposition (\ref{eq:Decomposition}):
Fix $n=2,3, \ldots $ and pick $\nu$ in the interval $(0, \nu_\star (\rho))$.  Thus, 
(\ref{eq:BoundOnRatio}) holds, and we have
\[
\left ( \frac{\mu(1)}{\nu} \cdot \frac{1 - \nu }{1-\mu(1)} \right )^{S_{L_n} -  L_n \nu } 
\leq 
\left ( \frac{\mu(1)}{\nu} \cdot \frac{1 - \nu }{1-\mu(1)} \right )^{(1-\nu)L_n}
\]
since $S_{L_n} \leq L_n$.
Using $\Gamma(0) < \Gamma(1)$ in (\ref{eq:Ratio}) we then conclude that
\[
\left ( \Gamma(1)^\nu \Gamma(0)^{1-\nu} \right )^{L_n}
\leq
\Gamma(1)^{S_{L_n}} \Gamma(0)^{L_n-S_{L_n}}
 \mbox{~on~}  [ S_{L_n} - L_n \nu > 0 ],
\]
whence
\[
\left ( 1 - \Gamma(1)^{S_{L_n}} \Gamma(0)^{L_n-S_{L_n}} \right )^{n-1} 
\leq 
\left ( 1 - \left ( \Gamma(1)^\nu \Gamma(0)^{1-\nu} \right )^{L_n}
\right )^{n-1} 
 \mbox{~on~}  [ S_{L_n} - L_n \nu > 0 ].
\]

Using these bounds in the definition of $E^+_n(\nu,L_n) $, we obtain
\begin{eqnarray}
\lefteqn{ E^+_n(\nu,L_n) } & &
\nonumber \\
&\leq&
\left ( 1 - \left ( \Gamma(1)^\nu \Gamma(0)^{1-\nu} \right )^{L_n}
\right )^{n-1} 
\cdot
\left ( \frac{\mu(1)}{\nu} \cdot \frac{1 - \nu }{1-\mu(1)} \right )^{(1-\nu)L_n}
\mathbb{P}_{\nu} 
\left [ S_{L_n} - \nu L_n > 0 \right ]
\nonumber \\
&\leq&
\left ( 1 - \left ( \Gamma(1)^\nu \Gamma(0)^{1-\nu} \right )^{L_n}
\right )^{n-1} 
\cdot
\left ( \frac{\mu(1)}{\nu} \cdot \frac{1 - \nu }{1-\mu(1)} \right )^{(1-\nu)L_n}.
\label{eq:BOUND+}
\end{eqnarray}

Next we turn to bounding $E^-_n(\nu,L_n) $. Because $\Gamma(0) < \Gamma(1) < 1$, we always have
\[
\left ( 1 - \Gamma(1)^{S_{L_n}} \Gamma(0)^{L_n-S_{L_n}} \right )^{n-1}  \leq 1 
\]
and exploiting the bound (\ref{eq:BoundOnRatio}) gives
\[
\left ( \frac{\mu(1) }{\nu} \cdot \frac{1-\nu}{1-\mu(1)} \right )^{ S_{L_n} - L_n \nu }  \leq 1
\mbox{~on~} [ S_{L_n} - L_n \nu \leq 0 ].
\]
We readily conclude
$E^-_n(\nu,L_n)  \leq  \mathbb{P}_{\nu}  \left [ S_{L_n} - L_n \nu \leq 0 \right  ]  \leq 1$
by applying these two bounds to the expression of $E^-_n(\nu,L_n)$.

Thus, in order to establish (\ref{eq:BoundedA}) we need only show that
\begin{equation}
\limsup_{n \rightarrow \infty} E^+_n ( \nu ,L_n ) < \infty
\label{eq:BoundedAA}
\end{equation}
for {\em some} $\nu$ in $(0, \nu_\star (\rho))$, possibly under additional conditions which
ensure that the constraint (\ref{eq:ConditionForNU+})
also holds. As per the discussion following Theorem \ref{thm:Zero-OneLaw+Part2}, 
the condition (\ref{eq:ConditionForNUstar+}) guarantees (\ref{eq:ConditionForNU+})
when $\nu$ is selected in the interval $(\alpha_+(\rho),\nu_\star (\rho) )$, as we do from now on.

First, let the sequence $\myvec{\rho}: \mathbb{N}_0 \rightarrow \mathbb{R}_+$ be
the unique sequence associated with the $\rho$-admissible scaling $\myvec{L}: \mathbb{N}_0 \rightarrow \mathbb{N}_0$.
For each $n=2,3, \ldots$ consider each of the factors in the bound at (\ref{eq:BOUND+}).
We find that
\begin{eqnarray}
\left ( 1 - \left ( \Gamma(1)^\nu \Gamma(0)^{1-\nu} \right )^{L_n}
\right )^{n-1} 
&=&
\left ( 1 - \left ( \Gamma(1)^\nu \Gamma(0)^{1-\nu} \right )^{\rho_n \ln  n}
\right )^{n-1} 
\nonumber \\
&\leq&
e^{-(n-1) \left ( \Gamma(1)^\nu \Gamma(0)^{1-\nu} \right )^{\rho_n \ln  n} }
\nonumber \\
&=&
e^{-\frac{(n-1)}{n} \cdot n^{ 1 + \rho_n \ln   \left ( \Gamma(1)^\nu \Gamma(0)^{1-\nu} \right ) } }
\label{eq:BOUND+A}
\end{eqnarray}
and
\begin{eqnarray}
\left ( \frac{\mu(1)}{\nu} \cdot \frac{1 - \nu }{1-\mu(1)} \right )^{(1-\nu)L_n}
&=&
\left ( \frac{\mu(1)}{\nu} \cdot \frac{1 - \nu }{1-\mu(1)} \right )^{(1-\nu)\rho_n \ln  n}
\nonumber \\
&=&
n^{ (1-\nu)\rho_n \ln  \left ( \frac{\mu(1)}{\nu} \cdot \frac{1 - \nu }{1-\mu(1)} \right ) }.
\label{eq:BOUND+B}
\end{eqnarray}

By the $\rho$-admissibility of the scaling ${\bf L}: \mathbb{N}_0 \rightarrow \mathbb{N}_0$, for every $\varepsilon > 0$
there exists a positive integer $n_\star(\varepsilon)$ such that
$\rho - \varepsilon < \rho_n < \rho + \varepsilon$ whenever $n \geq n_\star(\varepsilon)$.
On that range  the bounds (\ref{eq:BOUND+A}) and (\ref{eq:BOUND+B}) imply
\begin{eqnarray}
\left ( 1 - \left ( \Gamma(1)^\nu \Gamma(0)^{1-\nu} \right )^{L_n}
\right )^{n-1} 
\leq
e^{-\frac{(n-1)}{n} \cdot n^{ 1 + (\rho + \varepsilon)  \ln   \left ( \Gamma(1)^\nu \Gamma(0)^{1-\nu} \right ) } }
\label{eq:BOUND+AA}
\end{eqnarray}
and
\begin{eqnarray}
\left ( \frac{\mu(1)}{\nu} \cdot \frac{1 - \nu }{1-\mu(1)} \right )^{(1-\nu)L_n}
\leq
n^{ (1-\nu) ( \rho + \varepsilon)  \ln  \left ( \frac{\mu(1)}{\nu} \cdot \frac{1 - \nu }{1-\mu(1)} \right ) }
\label{eq:BOUND+BB}
\end{eqnarray}
as we recall that $\Gamma(0)$ and $\Gamma(1)$ both live in $(0,1)$ and the inequality (\ref{eq:BoundOnRatio}) holds.
Given that (\ref{eq:ConditionForNU+}) holds for the choice of $\nu$, then it is also the case that
\begin{equation}
1 + (\rho + \varepsilon)  \ln   \left ( \Gamma(1)^\nu \Gamma(0)^{1-\nu} \right )  > 0
\label{eq:ExtraCondition+Varepsilon}
\end{equation}
{\em provided} $\varepsilon > 0$ is selected small enough (as we do from now on).

Let $n$ go to infinity in (\ref{eq:BOUND+}). It is plain from (\ref{eq:BOUND+A}) that 
\[
\lim_{n \rightarrow \infty}
e^{-(n-1) \left ( \Gamma(1)^\nu \Gamma(0)^{1-\nu} \right )^{\rho_n \ln  n} } = 0
\]
by virtue of condition (\ref{eq:ExtraCondition+Varepsilon}), while (\ref{eq:BOUND+B}) implies
\[
\lim_{n \rightarrow \infty} 
\left ( \frac{\mu}{\nu} \cdot \frac{1-\nu}{1-\mu} \right )^{(1-\nu) \rho_n \ln  n } = \infty
\]
under (\ref{eq:BoundOnRatio}). Nevertheless, appealing to
the bounds (\ref{eq:BOUND+AA})  and (\ref{eq:BOUND+BB}), we have
$\lim_{n \rightarrow \infty } E^+_n(\nu,L_n) = 0$ in view of the fact that
\[
\lim_{n \rightarrow \infty} 
\left ( 
e^{-\frac{(n-1)}{n} \cdot n^{ 1 + (\rho + \varepsilon)  \ln   \left ( \Gamma(1)^\nu \Gamma(0)^{1-\nu} \right ) }  }
\cdot
n^{ (1-\nu) ( \rho + \varepsilon)  \ln  \left ( \frac{\mu(1)}{\nu} \cdot \frac{1 - \nu }{1-\mu(1)} \right ) }
\right )
= 0.
\]
This is because the first factor goes to zero like $e^{-n^\delta}$ (with $\delta > 0$)
while the second factor explodes to infinity like $n^\beta $ (with $\beta > 0$).
Obviously, $\limsup_{n \rightarrow \infty } E^-_n(\nu,L_n) \leq 1$
and the conclusion $\limsup_{n \rightarrow \infty } E_n(\nu,L_n) \leq 1$ follows.
This concludes the proof of the zero-law in Theorem \ref{thm:Zero-OneLaw+Part2}.
\myendpf

\section{Appendix: A proof of Proposition \ref{prop:Infinity-OneLaw+FirstMoment+Part2} -- The infinity-law}
\label{app:InfinityLawPart2}

Consider a $\rho$-admissible scaling
${\bf L}: \mathbb{N}_0 \rightarrow \mathbb{N}_0$ such that (\ref{eq:BasicCondition+Part2}) holds, or equivalently,
(\ref{eq:BasicCondition+Part2Bis}).
We already know that
\begin{equation}
1 + \rho \ln  G (\nu,\mu(1)) > 0 ,
\quad \nu \in (\nu_\star(\rho), \mu(1)),
\label{eq:Consequence1}
\end{equation}
and the convergence 
\[
\lim_{n \rightarrow \infty} n G(\nu,\mu(1))^{L_n}
= \infty,
\quad \nu \in (\nu_\star(\rho), \mu(1))
\]
follows by Lemma \ref{lem:TechnicalB} (with $C_n = G( \nu, \mu(1))$ for all $n=1,2, \ldots$).
By virtue of (\ref{eq:ChangeMeasure1}) the desired result 
$\lim_{n \rightarrow \infty} \bE{ I_n (L_n) } = \infty$
will be established if we show that
\begin{equation}
\liminf_{n \rightarrow \infty} E^+_n ( \nu ,L_n ) > 0
\label{eq:LimInf>0}
\end{equation}
for {\em some} $\nu$ in $(\nu_\star(\rho), \mu(1))$
possibly constrained by some additional condition.

Pick $\nu$ still in $(\nu_\star(\rho), \mu(1))$
for the time being, and fix $n=2,3, \ldots $. Because  (\ref{eq:BoundOnRatio}) holds here, we have
\begin{equation}
\left ( \frac{\mu(1)}{\nu} \cdot \frac{1 - \nu }{1-\mu(1)} \right )^{S_{L_n} -  L_n \nu } 
\geq 1
\mbox{~on~} [ S_{L_n} - L_n \nu > 0 ]
\end{equation}
so that
\begin{eqnarray}
E^+_n(\nu,L_n)
\geq
\mathbb{E}_{\nu} 
\left [
\left ( 1 - \Gamma(1)^{S_{L_n}} \Gamma(0)^{L_n-S_{L_n}} \right )^{n-1} 
\1{ S_{L_n} - \nu L_n > 0 }  \right ].
\label{eq:bound+}
\end{eqnarray}
Next, we write
\begin{eqnarray}
\left ( 1 - \Gamma(1)^{S_{L_n}} \Gamma(0)^{L_n-S_{L_n}} \right )^{n-1} 
=
\left ( 1 -  \left ( \Gamma(1)^{\frac{S_{L_n}}{L_n}} \Gamma(0)^{1-\frac{S_{L_n}}{L_n}} \right )^{L_n} \right )^{n-1} 
\label{eq:RewriteOfTerm}
\end{eqnarray}
and note that
\begin{equation}
\left |
\left ( 1 - \Gamma(1)^{S_{L_n}} \Gamma(0)^{L_n-S_{L_n}} \right )^{n-1} 
\right |
\leq 1 .
\nonumber
\end{equation}

Now further restrict the value of $\nu$ to the interval $(\nu_\star(\rho), \beta_-(\rho))$ 
discussed at the end of Section \ref{sec:Results}.
Condition  (\ref{eq:ConditionForNUstar-}) ensures that (\ref{eq:ConditionForNU-}) holds, and 
by Lemma \ref{lem:TechnicalA} (with $\nu_n = \frac{S_{L_n}}{L_n}$ for all $n=1,2, \ldots $,
with the help of (\ref{eq:RewriteOfTerm})), we have the convergence
\begin{equation}
\lim_{n \rightarrow \infty}
\left ( 1 - \Gamma(1)^{S_{L_n}} \Gamma(0)^{L_n-S_{L_n}} \right )^{n-1}
= 1 .
\quad \mathbb{P}_\nu-\mbox{a.s.},
\label{eq:ConvergenceAS}
\end{equation}
Indeed, the Strong Law of Large Numbers (under $\mathbb{P}_\nu$) yields the convergence
\[
\lim_{n \rightarrow \infty} \frac{S_{L_n}}{L_n} = \nu \quad \mathbb{P}_\nu-\mbox{a.s.},
\]
and this leads to the needed conclusion
\[
\lim_{n \rightarrow \infty} 
\left ( 1 + \rho_n \ln  \left ( \Gamma(1)^{\frac{S_{L_n}}{L_n}} \Gamma(0)^{1-\frac{S_{L_n}}{L_n}} \right )
\right )
= 
1 + \rho \ln  \left ( \Gamma(0)^{1-\nu} \Gamma(1)^\nu \right ) 
< 0
\quad \mathbb{P}_\nu-\mbox{a.s.}
\]
under (\ref{eq:ConditionForNU-}).

Pick $\varepsilon$ in $(0,1)$. It follows from the bound (\ref{eq:bound+}) that
\begin{eqnarray}
E^+_n(\nu,L_n)
\geq (1 - \varepsilon )
\mathbb{P}_{\nu} 
\left [ A_n(\varepsilon) \cap  [ S_{L_n} - \nu L_n > 0 ]  \right ],
\quad n=2,3, \ldots 
\label{eq:bound+Modified}
\end{eqnarray}
where for notational simplicity we have introduced the event
\[
A_n(\varepsilon)
= 
\left [
\left ( 1 - \Gamma(1)^{S_{L_n}} \Gamma(0)^{L_n-S_{L_n}} \right )^{n-1}  > 1 - \varepsilon 
\right ].
\]
Since a.s. convergence implies convergence in probability (under $\mathbb{P}_\nu$), it is plain from 
(\ref{eq:ConvergenceAS}) that
$\lim_{n \rightarrow \infty} \mathbb{P}_{\nu}  \left [ A_n(\varepsilon) \right ] = 1$.
On the other hand we also have
$
\lim_{ n \rightarrow \infty}  \mathbb{P}_\nu [  S_{L_n} - L_n \nu >  0  ] = \frac{1}{2}
$
by the Central Limit Theorem (under $\mathbb{P}_\nu$), whence
$
\lim_{n \rightarrow \infty} 
\mathbb{P}_{\nu} 
\left [ A_n(\varepsilon) \cap  [ S_{L_n} - \nu L_n > 0 ]  \right ] =  \frac{1}{2}
$
by standard arguments.
Therefore,
$\liminf_{n \rightarrow \infty}  E^+_n(\nu,L_n)  \geq (1 - \varepsilon)/2$ and the desired conclusion
$\liminf_{n \rightarrow \infty}  E^+_n(\nu,L_n) \geq 1$ follows since $\varepsilon $ is arbitrary in $(0,1)$.
This conclude the proof of the infinity-law in Proposition \ref{prop:Infinity-OneLaw+FirstMoment+Part2}.
\myendpf

\end{document}